\documentclass[twocolumn, twocolappendix]{aastex63}
\usepackage{amsmath,amstext}
\usepackage[T1]{fontenc}
\usepackage{apjfonts} 
\usepackage[figure,figure*]{hypcap}
\usepackage{commath}
\usepackage{color}

\usepackage{mhchem}
 
\newcommand{\eV}{\,{\rm eV}}
\renewcommand{\sec}{{\,\rm s}}

\newcommand{\kms}{\,{\rm km\,s^{-1}}}

\newcommand{\Msun}{{\rm \, M_{\odot}}}

\newcommand{\smetal}{Z_{\odot}}

\newcommand{\yr}{\, {\rm yr} }

\newcommand{\Myr}{\, {\rm Myr} }

\newcommand{\kpc}{\, {\rm kpc} }

\newcommand{\dd}{{\rm d}}

\newcommand{\cm}{{\, \rm cm}}

\newcommand{\K}{{\rm K}}
\newcommand{\erg}{{\rm \, erg}}

\ifx\arcsec\undefined
\newcommand{\arcsec}{\, {\rm arcsec}}
\fi

\renewcommand{\arraystretch}{1.4}

\newcommand{\eq}[1]{\begin{equation}#1\end{equation}}

\newcommand{\gathering}[1]{\begin{gather}#1\end{gather}}
\newcommand{\splitting}[1]{\begin{split}#1\end{split}}
\newcommand{\braket}[1]{\left(#1\right)}

\newcommand{\e}[1]{\times 10^{#1}}

\newcommand{\beqas}{\begin{eqnarray*}}
\newcommand{\eeqas}{\end{eqnarray*}}

\ifx\del\undefined
\newcommand{\del}[2]{\frac{\dd #1}{\dd #2}}
\fi

\ifx\appref\undefined
\newcommand{\appref}[1]{Appendix~\ref{#1}}
\fi
\newcommand{\fref}[1]{Figure~\ref{#1}}
\newcommand{\tref}[1]{Table~\ref{#1}}

\ifx\secref\undefined
\newcommand{\secref}[1]{Section~\ref{#1}}
\fi
\newcommand{\eqnref}[1]{Eq.(\ref{#1})}

\newcommand{\unit}[2]{\, {\rm #1}^{#2}}

\newcommand{\cs}{c_{\rm s}}

\ifx\dif\undefined
\newcommand{\dif}[2]{\frac{{\rm d}#1}{{\rm d}#2}}
\fi

\newcommand{\col}[1]{N_{\rm #1}}

\newcommand{\metal}{Z}
\makeatletter
\newcommand*{\rom}[1]{\expandafter\@slowromancap\romannumeral #1@}
\makeatother
\ifx\ion\undefined
\newcommand\ion[2]{#1$\;${\small\rmfamily\rom{#2}}\relax}
\fi

\newcommand{\CII}{\ion{C}{2}}

\newcommand{\OI}{\ion{O}{1}}

\newcommand{\abn}[1]{y_{\text{\rm #1}}}
\newcommand{\nh}{n_{\text{\rm H}}}

\newcommand{\nspe}[1]{n_{\text{\rm #1}}}
\newcommand{\partialdif}[2]{\dfrac{\partial#1}{\partial #2}}

\newcommand{\Kelvin}{{\rm \, K}}

\newcommand{\tcr}{t_{\rm cr}}

\ifx\fh\undefined
\newcommand{\fh}{f_{\rm h}}
\fi

\ifx\rmxaa\undefined
\newcommand{\rmxaa}{}
\fi

\newcommand{\rvir}{r_{\rm vir}}
\newcommand{\cn}{c_{\rm N}}

\newcommand{\potential}{\Phi}
\newcommand{\Tvir}{T_{\rm vir}}

   \newcommand{\sil}[4]{Z#1J#2M#3z#4}
   \newcommand{\simlabel}[4]{Z#1J#2M#3z#4}

   \newcommand{\zin}{z_{\scalebox{0.6}{\rm IN}}}

\shorttitle{Minihalo photoevaporation}
\shortauthors{Nakatani et al.}

\begin{document}

\title{Photoevaporation of Minihalos during Cosmic Reionization: Primordial and Metal-Enriched Halos}
\author[0000-0002-1803-0203]{Riouhei Nakatani}
\affiliation{RIKEN Cluster for Pioneering Research, 2-1 Hirosawa, Wako-shi, Saitama 351-0198, Japan}
\email{ryohei.nakatani@riken.jp}
\author[0000-0002-1369-633X]{Anastasia Fialkov}
\affiliation{Institute of Astronomy, University of Cambridge, Madingley Road, Cambridge CB3 0HA, UK}
\author[0000-0001-7925-238X]{Naoki Yoshida}
\affiliation{Department of Physics, School of Science, The University of
Tokyo, 7-3-1 Hongo, Bunkyo, Tokyo 113-0033, Japan}
\affiliation{Kavli Institute for the Physics and Mathematics of the Universe (WPI), UT Institute for Advanced Study, The University
of Tokyo, Kashiwa, Chiba 277-8583, Japan}
\affiliation{Research Center for the Early Universe (RESCEU), School of
Science, The University of Tokyo, 7-3-1 Hongo, Bunkyo, Tokyo 113-0033, Japan}

\begin{abstract}
The density distribution of the inter-galactic medium 
is an uncertain but highly important issue 
in the study of cosmic reionization.
It is expected that there are abundant gas clouds hosted by low-mass "minihalos" in the early universe, which act as photon sinks until photoevaporated by the emerging ultra-violet background (UVB) radiation.
We perform a suite of radiation hydrodynamics simulations to study the photoevaporation 
of minihalos. Our simulations follow hydrodynamics, non-equilibrium chemistry
and the associated cooling processes in a self-consistent manner. 
We conduct a parametric study by considering
a wide range of gas metallicity ($0\,\smetal \leq \metal \leq 10^{-3}\,\smetal$), halo mass ($10^3\Msun \leq M \leq 10^8\Msun$), UVB intensity ($0.01 \leq J_{21} \leq 1$), and turn-on redshift of ionizing sources ($10\leq \zin \leq 20$). 
We show that small halos are evaporated in a few tens million years, whereas larger mass halos survive for ten times longer. 
We show that the gas mass evolution of a minihalo can be characterized by 
a scaling parameter that is given by a combination of the halo mass, background radiation intensity, and redshift.
Efficient radiative cooling in metal-enriched halos 
induce fast condensation of the gas to form a dense, self-shielded core. 
The cold, dense core can become gravitationally unstable in halos with high metallicities.
Early metal enrichment may allow star formation in minihalos 
during cosmic reionization. 
\end{abstract}


\section{Introduction}

A broad range of observations including the cosmic microwave background radiation anisotropies have established
the so-called standard cosmological model in which the 
universe mainly consists of two unknown substances called dark energy and
dark matter.
According to the standard model,
small primeval density fluctuations generated in the very early universe grow by gravitational instability, to form 
nonlinear objects called dark matter halos.

Dark halos with mass $\sim 10^5$--$10^6\Msun$ are thought to be the birth place of the first generation of stars \citep{1997_Tegmark, 2003_YoshidaAbelHernquist}. 
Star-forming gas clouds are formed through condensation of the primordial gas by molecular hydrogen cooling. 
Population~III (\ion{Pop}{3}) stars are formed 
in the primordial gas clouds when the age of the universe is a few tens/hundreds million years 
\citep[e.g,][]{2002_AbelBryanNorman, 2006_Naoz, 2012_Fialkov}. 

Ultra-violet (UV) and X-ray radiation from the first stars and their remnants ionize and heat the inter-galactic medium (IGM), to initiate cosmic reionization.
Recent observations including
the measurement of the electron scattering optical depth \citep[e.g.,][]{2016_PlanckCollaboration}, the Gunn-Peterson trough in quasar spectra \citep{1965_GunnPeterson, Fan:2006, Banados:2018}, and Ly-$\alpha$ emission from star-forming galaxies \citep{Mason:2018} suggest that the process of reionization completes by $z\sim 6$ \citep[e.g.,][]{Weinberger:2020, Naidu:2020}. 
It is expected that early stages of reionization can be directly probed by future radio telescopes such as Square Kilometer Array
through observation of redshifted 21-cm emission from neutral hydrogen in the IGM \citep[see, e.g.,][for a recent review]{2016_Barkana, Mesinger:2019}. 

In the early phase of reionization, the density distribution of the IGM,
or the so-called gas clumping, is an important factor that critically sets the UV photon budget necessary for reionization.
Dense gas clouds hosted by cosmological minihalos can be significant 
photon sinks, and their existence and the abundance
affect the process and duration of reionization.
Unfortunately, it is nontrivial to derive the abundance 
of the gas clouds and to estimate the effective gas clumping factor,
because small gas clouds are effectively 
photo-evaporated by the emerging UV background radiation.
At the same time, stars can be formed in the gas clouds, 
which then act as UV photon {\it sources}. 

A number of studies have investigated early star formation under various environments \citep[e.g.,][]{1976_Low, 2000_Omukai, 2002_Schneider, 2003_Schneider, 2007_GloverJappsen, 2007_Jappsen, 2009_Jappsen, 2009_Jappsenb, 2009_Smith, 2018_Chiaki, 2019_ChiakiWise}.
Metal enrichment affects the evolution of gas clouds and subsequent star formation process 
through enhanced radiative cooling by heavy element atoms and dust grains \citep[e.g,][]{2005_Omukai, 2019_Hartwig}. 
Although details of low-metallicity star formation 
has been explored by recent numerical simulations \citep{2018_Chiaki},
the effect of metal-enrichment on halo photoevaporation has not been systematically studied. 
It is important to study the evolution of minihalos 
with a wide range of metallicities
in order to model the physical process of cosmic reionization
in a consistent manner.

Reionization begins as a local process in which an individual
radiation source generates an \ion{H}{2} region around it.
\cite{1986_Shapiro} and \cite{1987_ShapiroGiroux} 
use a one-dimensional model under spherical symmetry to study the ionization front (I-front) propagation through the IGM in a  cosmological 
context.
Radiative transfer calculations have been performed in a post-processing
manner by using the density field realized in cosmological simulations
\citep{1999_Abel, 1999_RazoumovScott, 2001_SokasianAbelHernquist, 2002_Cen, 2003_HayesNorman}. 
Fully coupled radiation-hydrodynamics simulations have been used to study reionization in a cosmological volume  \citep{2000_Gnedin, 2001_GnedinAbel, 2002_RicottiGnedinShull, 2004_SusaUmemuraa, 2004_SusaUmemurab, 2014_Wise, 2016_Xu}. Recent simulations of reionization explore the process of cosmic reionization employing large cosmological volumes of $\sim 100^3$ comoving Mpc$^3$ while resolving the first galaxies in halos with mass of $\sim 10^8 M_\odot$ 
\citep{Semelin:2017, Ocvirk:2018}. However, even these state-of-the-art simulations do not fully resolve minihalos 
nor are able to follow the process of photoevaporation,
and thus it still remains unclear 
how the small-scale gas clumping affects reionization.

\cite{2004_Shapiro} and \cite{2005_Iliev} study the dynamical evolution of minihalos irradiated by UV radiation.
They perform 2D radiation hydrodynamics simulations with  
including the relevant thermo-chemical processes. They explore a wide range of parameters such as halo mass, redshift, and the strength of UV radiation.
It is shown that gas clumping at sub-kiloparsec scales dominate absorption of ionizing photons during the early phase of reionization. 
An important question is whether or not the minihalos survive
under a strong UVB for a long time, over a significant fraction of the age of the universe.
Another interesting question is whether or not stars are formed in metal-enriched minihalos. If massive stars are formed, they also contribute to reionization and may thus imprint characteristic features in the $21\cm$ signals \citep{2016_Cohen}.

In the present paper, we perform a large set of high-resolution radiation hydrodynamics simulations of 
minihalo photoevaporation. 
We aim at investigating systematically the effects of metal enrichment on the photoevaporation.
We evaluate the characteristic photoevaporation time and study its metallicity dependence. We also develop an analytic model
of photoevaporation and compare the model prediction with our
simulation result.
The rest of the present paper is organized as follows.
In \secref{sec:methods}, we explain the details of our computational methods. 
The simulation results are presented in \secref{sec:results}. 
We develop an analytical model 
that describes the physical process of minihalo photoevaporation in \secref{sec:analytic}. 
We discuss the physics of minihalo photoevaporation in \secref{sec:discussion}. 
Finally, we give summary and concluding remarks in \secref{sec:conclusions}.

Throughout the present paper, we assume a flat $\Lambda$CDM cosmology
with $(\Omega_{\rm m}, \Omega_\Lambda, \Omega_{\rm b}, h) = (0.27, 0.73, 0.046, 0.7)$ \citep{2011_Komatsu}.
All the physical quantities in the following are given in physical units.

\section{Numerical Simulations} \label{sec:methods}
	We perform 
	radiation-hydrodynamics simulations 
	of minihalo photoevaporation by an external UV background radiation. 
	 We run a set of simulations systematically by varying the gas metallicity, dark matter halo mass, intensity of the radiation background and redshift. 
	
\begin{figure*}[htbp]
		\begin{center}
		\includegraphics[clip, width = \linewidth-5cm]{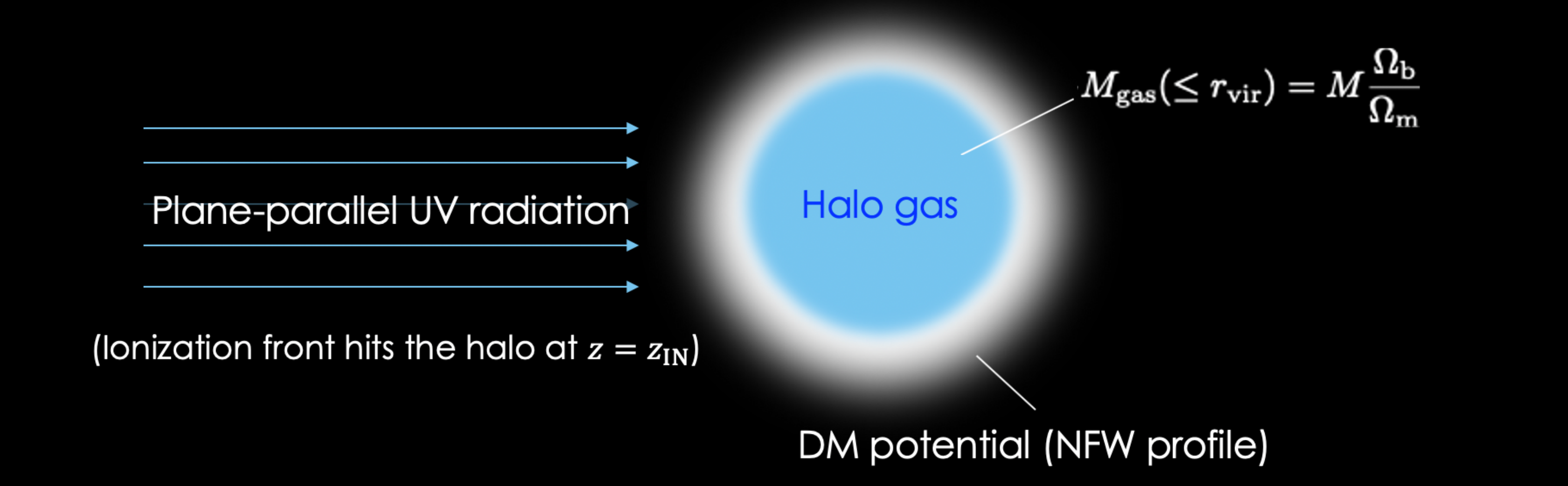}
		\caption{
			We consider plane-parallel radiation incident on a halo with mass $M$ and metallicity $\metal$.
			The I-front reaches the halo at $z = \zin$. 
			}
		\label{fig:schematic}
		\end{center}		
	\end{figure*}

	Our simulation set up is schematically shown in \fref{fig:schematic}. 
	
	The numerical method is essentially the same as in \cite{2019_Nakatani}, where we study
	the dynamical evolution of molecular gas clouds 
	exposed to an external UV radiation field.
	Briefly, we use a modified version of PLUTO \citep[version 4.1;][]{2007_Mignone}
	that incorporates ray-tracing radiative transfer
	of UV photons 
	and non-equilibrium chemistry.
	The details of the implemented physical processes are found in  
	\cite{2018_Nakatani, 2018_Nakatanib}.
	
	The simulations are configured with 2D cylindrical coordinates. 
	The governing equations are
	\gathering{
		\frac{\partial \rho_{\rm b}}{\partial t} + \nabla \cdot \rho_{\rm b} \vec{v} =  0 ,	
		\label{eq:continuity}\\
		\frac{\partial \rho_{\rm b} v_R}{\partial t} + \nabla \cdot \left( \rho_{\rm b} v_R \vec{v} \right)              
		=  -\frac{\partial P}{\partial R}  - \rho_{\rm b} \partialdif{\potential}{R},
			\\
		\frac{\partial \rho_{\rm b} v_x}{\partial t} + \nabla \cdot \left( \rho_{\rm b} v_x \vec{v} \right)    
		= - \frac{\partial P}{\partial x } -\rho_{\rm b} \partialdif{\potential}{x} , \\
		\frac{\partial \rho_{\rm b} E}{\partial t} + \nabla \cdot \left(\rho_{\rm b} H \vec{v} \right) 
		= - \rho_{\rm b} \vec{v} \cdot \nabla \potential  
		+\rho_{\rm b} \left( \Gamma -\Lambda    \right),                        \\
		\frac{\partial \nh y_i }{\partial t} + \nabla \cdot \left( \nh y_i \vec{v} \right)
		= \nh C_i       ,                       \label{eq:chemevoeq}
	}
	where $R$ and $x$ are the physical radial distance and vertical height, respectively,
	and $t$ is proper time in the frame of the halo. 
	We denote the gas density, velocity, and pressure as $\rho_{\rm b}$, $\vec{v} = (v_R, v_x)$, 
	and $P$. In the equation of motion, $\potential$ is the external gravitational potential of the host dark halo.
	In the energy equation, $E$ and $H$ are the total specific energy  
	and total specific enthalpy, and
	$\Gamma$ and $\Lambda$ are the specific heating rate 
	and specific cooling rate.
	The abundance of $i$-th chemical species, $\abn{i}$, is
	defined by the ratio of the species' number density $n_i$ 
	to the hydrogen nuclei number density $\nh$.
	The total reaction rate is denoted as $C_i$. The gas is composed of eight chemical species:
	H, \ce{H+}, \ce{H2}, \ce{H2+}, He, CO, \ce{C+}, O, and \ce{e-}.
	The elemental abundances of carbon and oxygen are normalized by
	the assumed metallicity $\metal$ as
 	$0.926\e{-4} \, \metal/\smetal$ 
	and $3.568\e{-4} \, \metal / \smetal$, respectively \citep{1994_Pollack, 2000_Omukai}.

	We consider halos with a wide range of masses of 
	$10^{3} \Msun \leq M \leq 10^8\Msun$. Each halo
	has a Navarro, Frenk \& White density profile scaled appropriately \citep{1997_NFW}.
	We assume that the halo potential is fixed and is given by a function of 
	the spherical radius $r \equiv \sqrt{R^2+x^2}$ as
	\eq{
		\rho_{\rm DM} \propto \frac{1 }{\cn \xi ( 1 + \cn \xi)^2},	\label{eq:nfw}
	}
	where $\cn$ is the concentration parameter, 
	and $x$ is the spherical radius normalized by the virial radius, i.e. $\xi \equiv r / r_{\rm vir}$.	
	The virial radius of a halo collapsing at redshift $z$ is
	\eq{
	\splitting{
		r_{\rm vir} = &1.51 \braket{\frac{\Omega_m h^2}{0.141}}^{-1/3}
							\braket{\frac{M}{10^8 \,\Msun}}^{1/3} \\
						&\times	\braket{\frac{\Delta_{\rm c}}{18\pi^2}}^{-1/3}
							\braket{\frac{1+z}{10}}^{-1}\,\kpc	\label{eq:rvir}
							}
	}
	where $\Delta_{\rm c}$ is the overdensity relative to the critical density 
	of the universe at the epoch.
	We adopt $\Delta_{\rm c}  =18\pi^2$.
	Then the halo potential $\potential$ is explicitly given by
	\gathering{
		\potential (r) 
					=  -  V_{\rm c}^2
					\frac{\ln\braket{1 + \cn \xi}}{\cn \xi }
					\frac{\cn }
					{\ln(1+\cn) - \dfrac{\cn}{1+ \cn}},
					\label{eq:potential}
	\\
	V_{\rm c} \equiv \sqrt{\frac{GM}{\rvir}}
	}

	The initial conditions are given by assuming a fully atomic, isothermal gas in hydrostatic equilibrium with the virial temperature
	\eq{
	\Tvir	=	\frac{GM \mu m_{\rm p}}{2\rvir k_{\rm B}},
	}
	where $\mu$ is mean molecular weight, 
	$m_{\rm p}$ is the proton mass,
	and $k_{\rm B}$ is the Boltzmann constant. 
	The initial density profile is 
	\eq{
		\rho_{\rm b} (r)
			=  \hat{\rho_{\rm b}} \exp\left[-\frac{\potential }{ k_{\rm B}T_{\rm vir}/ \mu m_{\rm p}}  \right], 	
			\label{eq:densityprofile}
	}
	where $\hat{\rho_{\rm b}}$ is the normalization factor 
	\eq{
			\hat{\rho_{\rm b}} \equiv 
			\frac{M 
			\Omega_{\rm b} \Omega_{\rm m}^{-1}}
			{
			\displaystyle
			\int _0 ^{\rvir} \dd r\, 4\pi r^2
			\exp\left[-\dfrac{\potential }{ k_{\rm B}T_{\rm vir}/ \mu m_{\rm p}}  \right]}.
	}
	With this normalization, the mass ratio of baryons to dark matter within $\rvir$ equals the global cosmic baryon fraction
	$f_{\rm b} = \Omega_{\rm b} / \Omega_{\rm m}$.
	The initial density profile is specified by $M$, $\cn$ and $\xi$.
	Note that $\hat{\rho_{\rm b}}$ is not the central density
	but is a geometry-weighted average density,
	which is independent of $M$ but scales as $\propto (1+z)^3$
	for fixed $\Delta_{\rm c}$ and $\cn$. 

	The initially fully atomic gas in the halo is exposed to  plane-parallel UV radiation as illustrated in Figure \ref{fig:schematic}.
	We follow photoionization by extreme UV (EUV; $13.6\eV < h\nu < 100\eV$) photons
	and photodissociation by far-UV photons in the Lyman-Werner (LW) band (FUV; $11.2\eV \lesssim h\nu < 13.6 \eV$).
	The UV spectrum is given by 
	\eq{
		J(\nu) = J_{21} \braket{\frac{\nu}{\nu_1}}^{-\alpha} \e{-21} 
		\unit{erg}{}\unit{s}{-1}\unit{cm}{-2}\unit{Hz}{-1}\unit{sr}{-1},	\label{eq:UVSED}
	}
	where $\nu_1$ is the Lyman limit frequency (i.e., $h\nu_1 = 13.6\eV$). 
	We set the UV spectral slope $\alpha = 1$ and 
    consider $J_{21}$ in the range $0.01 \leq J_{21} \leq 1$ \citep{1996_ThoulWeinberg}. 
	We calculate the photodissociation rate with
	taking into account the self-shielding of hydrogen molecules
	\citep{1996_DraineBertoldi, 1996_Lee}.

	Heating and cooling rates are calculated self-consistently with the
	non-equilibrium chemistry model of \cite{2019_Nakatani}.	
	Major processes are photoionization heating,
	Ly{\rm $\alpha$} cooling,
	radiative recombination cooling, 
	\CII~line cooling, 
	\OI~line cooling, 
	\ce{H2} line cooling,
	and CO line cooling. 
	The corresponding 
	heating/cooling rates are found in
	\cite{2018_Nakatani, 2018_Nakatanib}.
	FUV-induced photoelectric heating is not effective with FUV intensity and metallicity of interest in this study. We omit it from our thermochemistry model. 
	For the present study, we also implement the Compton cooling by the CMB photons interacting with free electrons with physical density $n_e$ as 
	\eq{
		\Lambda_{\rm Comp} = 5.65\e{-36}\nspe{e} (1+z)^4 (T - T_{\rm CMB}) \unit{erg}{}\cm^{-3} \unit{s}{-1},
	}
	where $T_{\rm CMB}$ is the CMB temperature given by $T_{\rm CMB} = 2.73(1+z)\Kelvin$.

	Our computational domain extends $0\kpc \leq R \leq \rvir$ and $-\rvir  \leq x \leq \rvir$.  
	We define computational grids uniformly spaced with the number of grids $N_R \times N_x = 320\times 640$. 	
	UV photons are injected from the boundary plane at $x = -\rvir$. 
	We assume that the cosmological I-front arrives at the plane at a redshift of $\zin$. 	
	All our runs start at this time denoted as $t = 0 \yr$. 
	Note that the external halo potential is fixed; we do not consider growth of halo mass.
	We discuss potential influences of this simplification in \secref{sec:discussion}. 
	
	We run a number of simulations with varying three parameters in the range
	$0\, \smetal \leq \metal \leq 10^{-3} \, \smetal$, $0.01 \leq J_{21} \leq 1$,
	$10^{3} \Msun \leq M \leq 10^8 \Msun$,
	and $10\leq \zin \leq 20$, respectively. 
	A total of 495 ($=5\times3\times11\times3$) 
	simulations are performed.
	Hereafter, we dub each run based on the assumed values of the parameters. A simulation with $(\metal, J_{21}, M, \zin) = (10^{-a}\smetal, 10^{-b}, 10^{c}\Msun, d)$
	is referred to as ``\sil{$a$}{$b$}{$c$}{$d$}''.
	For example, \sil{$\infty$}{0}{5.5}{15} indicates 
	$(\metal, J_{21}, M, \zin) =(0 \,\smetal, 1, 10^{5.5}\Msun, 15)$.

\section{Photoevaporation Process} \label{sec:analytic}
\begin{figure*}
    \centering
    \includegraphics[clip, width = \linewidth]{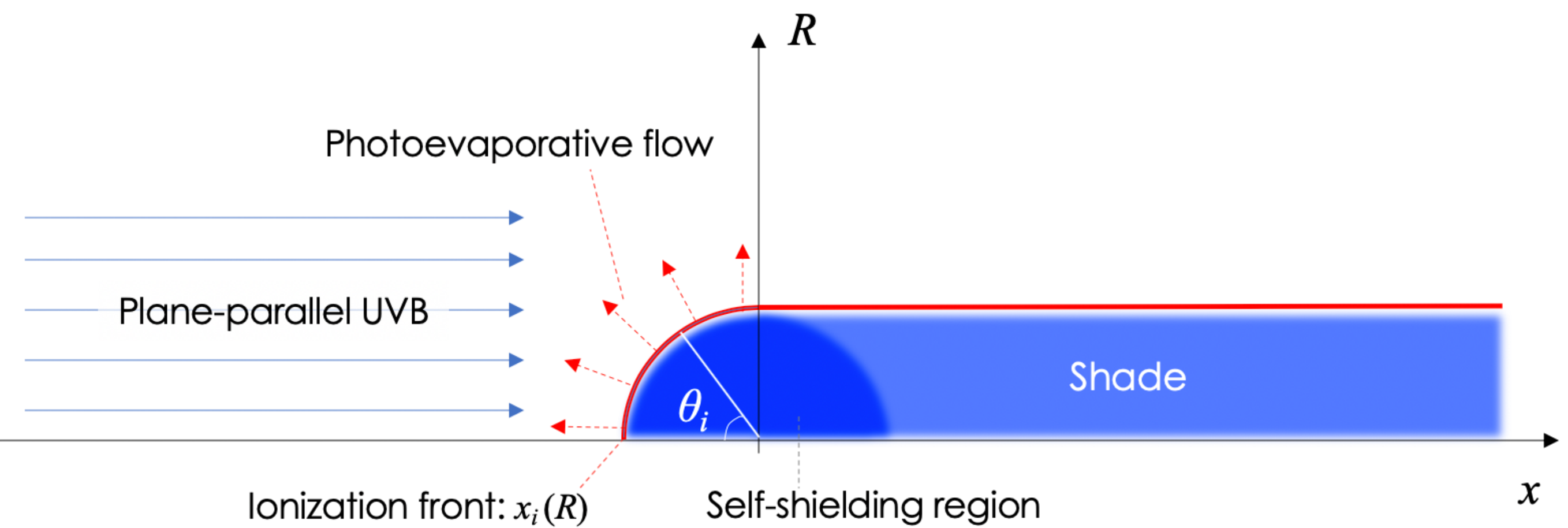}
    \caption{Schematic view of our analytic model in \secref{sec:analytic}. Planer UVB is incident on the halo, dividing it into self-shielded region (represented by the blue hemisphere) and photoevaporative flow region. Photoevaporative flows launch from the outermost layer of the self-shielded region. The red line indicates the locus of the I-front, $x_i(R)$. There is a UVB-shade in the downstream region, where the gas is neutral.}
    \label{fig:analytic}
\end{figure*}
Before presenting the results from a number of simulations, it would be
illustrative to describe the basic physics of photoevaporation.
It also helps our understanding of the overall evolution of a UV-irradiated minihalo. To this end, we develop an analytic model and 
evaluate the photoevaporation rate numerically so that it can be compared directly with our simulation results.

Photoevaporation is driven by gas heating associated with
photo-ionization.
Incident UV radiation ionizes the halo gas, forming a sharp boundary between the ionized and neutral (self-shielded) regions.
Photoevaporative flows are launched from the outermost layer of the self-shielded region. 
The number of UV photons incident on the self-shielded region is equal to that of evaporating gas particles. 
The gas mass evolution of a halo can be described as 
\eq{
\splitting{
	\frac{\dd M _{\rm s}}{\dd t} & =  - m \int_{\partial V_{\rm s}} \dd \vec{S}_i  \cdot \hat{x} \, J \\
	&=	- m \int_{V_{\rm s}} \dd V \, \nabla \cdot \hat{x} \,J \\
	&	=	 - m \int _0^{R_i } 2\pi R \, J (R, x_i) \, \dd R, 
	}\label{eq:masslosseq}
}
where $M_{\rm s}$ is the total gas mass in the self-shielded region, $m \nh \equiv \rho_{\rm b}$, $V_{\rm s}$ is the volume of the self-shielded region, $\hat{x}$ is a unit vector in the $x$-direction, $J$ is total photon number flux, $R_i$ is the maximum radial extent of the self-shielded region (i.e., the radial position of the I-front), and $x_i(R)$ gives the locus of the I-front on the $R$--$x$ plane (\fref{fig:analytic}).

We define the following dimensionless quantities:
$\vec{\xi} = (\xi_x, \xi_R) \equiv (x/\rvir, R/\rvir)$, 
$\tilde{t} = t/ (\rvir V_{\rm c}^{-1}) $, 
$\tilde{M} = M_{\rm s} / (f_{\rm b} M )$ with $f_{\rm b} = \Omega_{\rm b} / \Omega_{\rm m}$,
and $\tilde{J} = J / J_0$ $(J_0\equiv 8.1\e{5}\, J_{21} \unit{s}{-1}\unit{cm}{-2})$; 
and rewrite \eqnref{eq:masslosseq} in a dimensionless form
\eq{
	\frac{\dd \tilde{M} _{\rm s}}{\dd \tilde{t} } = 
	- \frac{J_0 \rvir^3 m }{f_{\rm b} M V_{\rm c}}
	\int _0^{\xi_i } 2\pi \xi_R \, \tilde{J} (\xi_R, \xi_{x, i}) \, \dd \xi_R. 
	\label{eq:nondimmassloss}
}
The ionizing photon number flux at the I-front, $J(R, x_i)$, is equal to $J_0$ minus the total recombinations along the ray up to the I-front. Note that $J(R,x_i)$ depends on the recombination coefficient and the density, and also on the velocity profile of the photoevaporative flows. 

We approximate the I-front to be a hemispheric surface facing toward the incident radiation in the region $x_i < 0$,
i.e., $x_i = - R_i \sqrt{1 - R^2/ R_i^2}$.
In the other region $x_i \geq 0$, the I-front lies at the surface of a cylinder with radius $ R_i $. 
We further assume that photoevaporative flows are spherically symmetric in $x_i < 0$ and the ionized gas is isothermal with $T = 10^4\Kelvin$.
The wind velocity is assumed to be the sound speed of the ionized gas, $c_i$. 
The ionizing photon number flux at the I-front is then given by
\gathering{
	{J}(R, x_i) = \frac{2 J_0}{1 + \sqrt{1 + 4\dfrac{\alpha_{\rm B} R_i J_0}{c_i^2} \dfrac{\theta_{R,i}}{\sin \theta_{R,i} }} } \nonumber \\
	\theta_{R,i} \equiv \arccos\sqrt{1 - \braket{\frac{R}{R_i}}^2}
	= \arccos\sqrt{1 - \braket{\frac{\xi_R}{\xi_i}}^2}, \nonumber 
}
where $\alpha_{\rm B}$ is the case-B recombination coefficient. 
With these results, \eqnref{eq:nondimmassloss} reduces to 
\gathering{
	4\pi \xi_i^2 \tilde{\rho_{\rm b}}(\xi_i) \frac{\dd \xi_i}{\dd \tilde{t} } = 
	- \eta
	\int _0^{\xi_i }  \frac{4\pi \xi_R \, \dd \xi_R}{1 + \sqrt{1 + 4q\xi_i 
	\dfrac{\theta_{R,i}}{\sin \theta_{R,i}}
	} 
	}
	\label{eq:difmassloss}\\ 
	\eta \equiv \frac{J_0 \rvir^3 m }{f_{\rm b}  M V_{\rm c}}
	\approx 14 J_{21} \braket{\frac{M}{10^6\Msun}}^{-1/3} \braket{\frac{1+z}{11}}^{-7/2}, \nonumber \\
	q \equiv \dfrac{\alpha_{\rm B} \rvir  J_0}{c_i^2} 
	\approx 1.7\e{2} J_{21} \braket{\frac{M}{10^6\Msun}}^{1/3} \braket{\frac{1+z}{11}}^{-1}. \nonumber 
}
Note that the dimensionless parameter $\eta$ effectively measures the ratio of Hubble time to the ionization time scale.
The other parameter $q$ quantifies the magnitude of UV absorption in the photoevaporative flows; absorption is negligible if $q \ll 1$, while it is significant if $q \gg 1$. 
Although the differential equation is not solved analytically, we can derive the asymptotic behaviour of the gas mass in a few limiting cases. 
For $q \gg 1$, the right-hand-side coefficient is approximately proportional to $\simeq \eta q^{- 1/2} \propto J_{21}^{1/2} M^{-1/2} (1+z)^{-3}$.
It follows that there is a similarity in the gas mass evolution among halos having similar $\eta q^{-1/2}$ and initial $\xi_i$.
Hence we define the similarity parameter as 
\eq{ 
    \chi \equiv \eta q ^ {-1/2}.    \label{eq:chi}
}
The initial $\xi_i$ is determined by the radius 
at which the I-front turns to the D-critical type from R-type \citep[e.g.,][]{1978_Spitzer, 1989_Bertoldi}.
The radius is obtained by numerically solving the integral equation
\gathering{
	\int_{-\infty}^{-\xi_i} \tilde{\rho_{\rm b}}^2 (\xi_R, \xi_x) \,\dd\xi_x = 
	\frac{J_0 }{ \rvir n_0^2 \alpha_{\rm B}} 
	\braket{1 
	- \frac{2c_i  n_0 }{  J_0} \tilde{\rho_{\rm b}} (\xi_i)
	} \nonumber \\
	n_0 \equiv \frac{f_{\rm b} M} {m \rvir^3}  \nonumber \\
	\tilde{L}_{\rm s}\equiv \frac{J_0 }{ \rvir n_0^2 \alpha_{\rm B}} \nonumber \\
	\tilde{u}_{\rm IF} \equiv \frac{2c_i  n_0 }{  J_0}, \nonumber 
}
with the initial density profile of \eqnref{eq:densityprofile}. 
Typically, the initial $x_i$ is larger
for lower $J_0$ and for higher $n_0$ (i.e., higher $\zin$). 
We list $q, \eta, q, \chi, \tilde{L}_{\rm s}, $ and $\tilde{u}_{\rm IF}$ for each of our runs in \tref{tab:data} of \appref{sec:supplymental}. 

In the above model, we have assumed a constant flow velocity, 
but in practice the flow is accelerated within the ionized boundary layer after 
launched with a small, negligible velocity. Also, we do not consider gravitational force by the host halo, which can decelerate the photoevaporative flows. In \secref{sec:similarities}, we will introduce a few corrections 
to these simplifications and
examine carefully the similarity of gas mass evolution by comparing with the simulation results.

\section{Simulation Results} \label{sec:results}
We first describe the dynamical evolution of 
a minihalo in our fiducial case, and examine the effect of metal- and dust-cooling in \secref{sec:result1}. 
Then, we focus on the photoevaporation rates 
and study the dependence on metallicity and on 
halo mass in \secref{sec:massloss}. The dependence of the photoevaporation 
rates on radiation intensity and the turn-on redshift is
studied in \secref{sec:result2} and \secref{sec:result3}.
We summarize the halo mass evolution in \secref{sec:similarities}. 
Then we provide an analytical fit to the derived mass evolution as a function of time in \secref{sec:evatime}. 
For convenience, we term halos with $\Tvir > 10^4\Kelvin$ atomic cooling (massive) halos in the following sections. The corresponding mass range is $M \gtrsim 10^{7.5}\Msun$ ($ 10^7\Msun$) at $z = 10$ (20). Lower-mass halos ($\Tvir < 10^4\Kelvin$) are referred to as low-mass halos; those with $M \gtrsim 10^{6.5}$--$10^7\Msun$ ($10^6$--$10^{6.5}\Msun$) at $\zin = 10$ (15--20) are specifically called molecular cooling halos. 

\subsection{Photoevaporation and Metallicity Dependence}	\label{sec:result1}
\begin{figure*}[h]
\begin{center}
\includegraphics[clip, width = \linewidth]{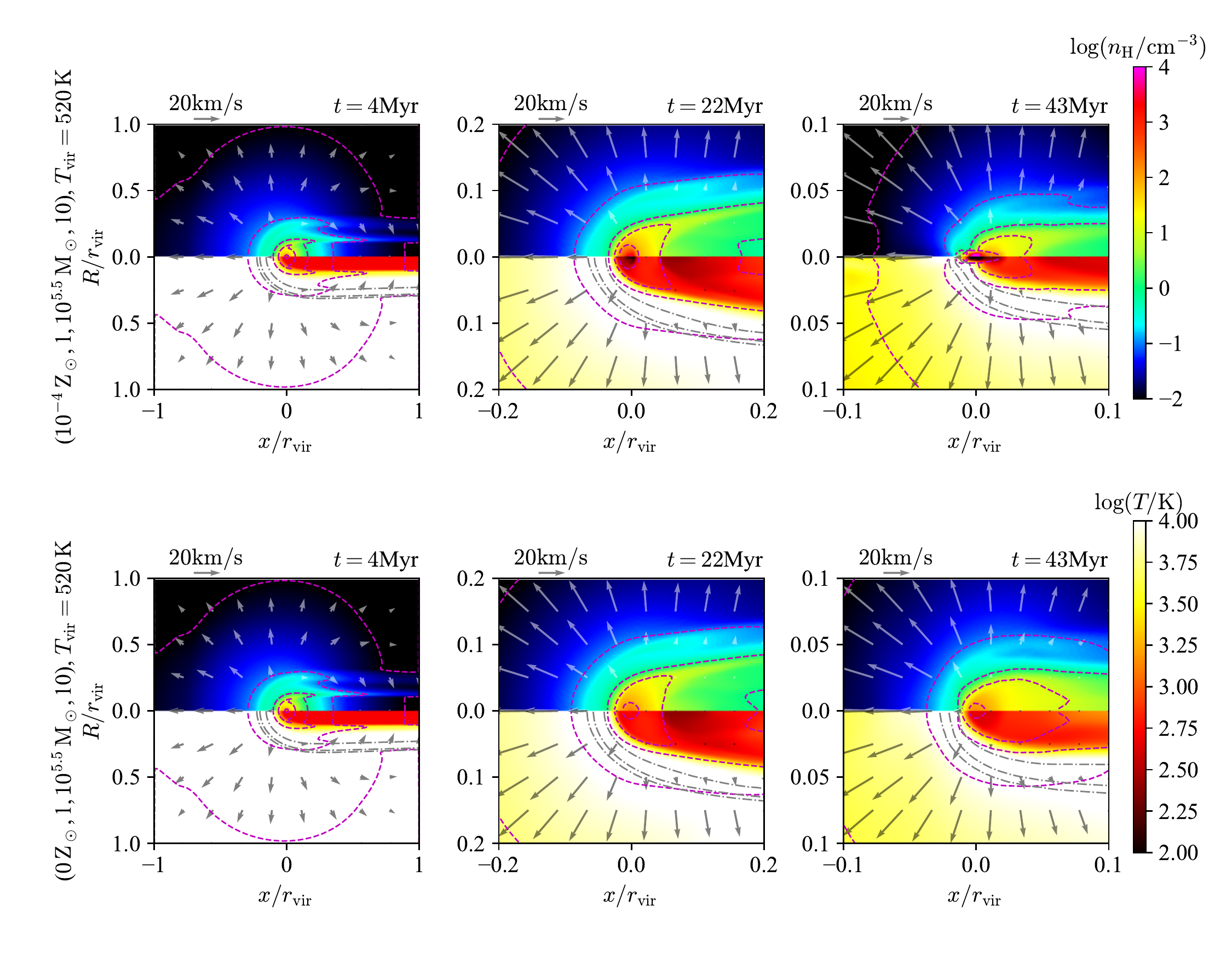}
\caption{
    Snapshots of \sil{4}{0}{5.5}{10} (top panels) and \sil{$\infty$}{0}{5.5}{10} (bottom panels).
	The upper and lower half of each panel indicate the density and temperature distributions, respectively. 
	The color bars are shown at the right. 
	The magenta dashed lines are density contours for $\nh = 10^{-2},10^{-1},1, 10, 100 \cm^{-3}$, 
	and the gray dot-dashed lines are ionization degree for $0.1, 0.5, 0.9$. 
	The arrows represent velocity fields and are scaled by the magnitude. 
	The reference arrow length for $20\kms$ is shown at the top left in each panel. 
	Note that the view closes-up as time goes for clarity. 
	The planer UV field is incident on the computational 
	domain at $x/\rvir = -1$. 
	The UV-heated gas has a temperature of $\sim 10^4\Kelvin$ 
	and streams off the halo. 
	The self-shielded regions (orange regions in the temperature maps) have relatively low temperatures. 
}
\label{fig:snapshots}
\end{center}
\end{figure*}

\fref{fig:snapshots} shows the density and temperature distributions for a halo with $M=10^{5.5}\Msun$ irradiated by the UV background with $J_{21} = 1$ at $\zin =10$.
We compare the results with two different metallicities of $\metal =0\,\smetal$ and $\metal=10^{-4}\,\smetal$ (i.e., \sil{$\infty$}{1}{5.5}{10} and \sil{4}{1}{5.5}{10}). 
In both runs, we find hot, ionized gas flows (hereafter ``wind region'') 
and a cold, dense region (hereafter ``self-shielded region''). 
The boundary between the two is the launching "base" of the photoevaporative flows. 
In \fref{fig:snapshots}, the base appears as a transitional layer that divides 
a hot ($\sim 10^4\Kelvin$; white) region and a cool ($\lesssim 10^3\Kelvin$; orange) region.
The wind regions are heated by EUV photons, 
and the temperature is $\sim 10^4\Kelvin$ near the base,
but quickly decreases as the wind expands.

The self-shielded region of a 
metal-free halo contracts slowly
because of inefficient cooling.
With a slight amount of heavy elements, atomic cooling such as \CII{} and \OI{} cooling becomes effective, 
and also the grain-catalyzed \ce{H2} formation reaction produces abundant \ce{H2} molecules ($y_{\ce{H2}} \sim 10^{-4}$)
that further enhance cooling efficiency. 
Since the thermal coupling of dust and gas is weak at low densities, 
the dust temperature does not become high enough to sublimate.
The radiative cooling rates are roughly comparable
for \ce{H2}, \CII{}, and \OI{} species 
in metal-rich halos ($\metal \gtrsim 10^{-4}\,\smetal$).
Typically, the most efficient coolant is
\ce{H2} in a large portion of the self-shielded region,
whereas \CII{} cooling is dominant only in the central, very dense region ($\gtrsim10^{3}\cm^{-3}$).
The central low temperature ($T \sim 10^2\Kelvin$) part in the upper panels of \fref{fig:snapshots} is formed through the 
efficient atomic cooling. 


The self-shielded region of the metal-free halo (\sil{$\infty$}{1}{5.5}{10}) has a temperature close to the
virial temperature of the halo.
The wind region has similar thermal and chemical structure 
in the two runs shown in \fref{fig:snapshots}. Lyman~${\rm \alpha}$ cooling 
is the dominant cooling process near the I-front, whereas Compton cooling is 
important in outer regions. Note that the efficiency of the latter depends on the cosmological redshift $\zin$. The gas temperature in the wind region is $\sim 5000$--$10000 \Kelvin$ and decreases as the gas expands outward.

Since the cooling time is progressively shorter in the central, denser part, 
the gas cools and condenses in an inside-out manner.
A dense core forms quickly at the halo center, as can be seen in the time evolution
in \fref{fig:snapshots}.
In metal-enriched halos, a sufficient amount of \ce{H2} molecules is
formed via grain-catalyzed reactions, even though the incident radiation continuously dissociates \ce{H2}.
With increasing metallicity, \ce{H2} molecules form more rapidly,
and \ion{C}{2} and \ion{O}{1} line cooling also lower the gas temperature. 
An important effect of metal cooling is to lower the minimum 
mass for gas cloud collapse. 
We discuss whether or not star formation takes place in low-mass, low-metallicity halos in \secref{sec:starformation}. 

We have focused on the results with $\zin = 10$
and with the fiducial UV radiation intensity 
$J = 10^{-21} \erg \sec^{-1} \cm^{-2} {\rm Hz}^{-1} {\rm sr}^{-1}$.
Essentially the same physical processes operate  
in other cases. 
With $\sim 10^2$--$10^3$ times higher LW intensities, \ce{H2} molecules are almost completely photodissociated in the halo \citep{2010_ShangBryanHaiman, 2012_Agarwal, 2014_ReganJohanssonWise, 2015_Hartwig, 2016_ReganJohanssonWisea, 2017_Schauer}.
We note that primordial gas clouds formed under strong LW radiation are
suggested to be possible birth places of massive black holes \citep{2001_Omukai, 2010_ShangBryanHaiman, 2012_Agarwal, 2014_ReganJohanssonWise, 2015_Hartwig, 2016_ReganJohanssonWisea, 2017_Schauer}. 
In metal-enriched halos, 
the gas can still cool and condense by metal and dust cooling
even under strong UV radiation.

In contrast to low-mass, \ce{H2}-cooling halos, Ly$\alpha$ cooling dominates in 
halos with $\Tvir \gtrsim 20000\Kelvin$.
Since the efficiency of Ly$\alpha$ cooling is independent of metallicity, 
the gas condensates quickly in the massive halos with $M \gtrsim 10^{7.5}\Msun$. 
If the gas is enriched to  $\metal \gtrsim 10^{-3}\,\smetal$,
dust-gas collisional heat transfer 
causes efficient gas cooling for $\nh \gtrsim 10^7\cm^{-3}$ \citep{2005_Omukai, 2008_Omukai, 2016_Chiaki}. 


Our findings are largely consistent with \cite{2014_Wise} regarding
the correspondence between halo mass and dominant cooling processes.
Halos with $M \gtrsim 10^7\Msun$ are expected to be 
metal-enriched by Pop~III supernovae triggered in the progenitor halo or in nearby halos,
and thus metal cooling is dominant or comparable to atomic/molecular hydrogen cooling. 
Following \cite{2014_Wise}, we call such halos  ``metal-cooling halos'', which have masses between the atomic cooling limit ($\sim 10^{7.5}$--$10^8\Msun$) and the upper mass limit of molecular cooling halos ($\sim 10^{6.5}$--$10^7\Msun$). We find similarly efficient metal cooling for $M \gtrsim 10^7\Msun$, but the most important effect of metal enrichment in our cases is to lower the molecular cooling limit by allowing formation of \ce{H2} through grain-catalyzed reactions, especially at $\metal \gtrsim10^{-4}\,\smetal$. 

The gas in massive halos ($\Tvir > 10^4\Kelvin$) is gravitationally bound even under strong UV radiation.
We find rather small mass loss in the runs with $M = 10^{7.5}\Msun$. 
Approximately 10\% of the initial gas mass is lost via photoevaporation, 
but the diffuse, ionized gas follows the concentrated gas toward the center. 
This process slightly recovers the total gas mass within $\rvir$. 
For halos with $\Tvir > 10^4\Kelvin$, 
outgoing flows are not excited, and all of the baryons concentrate 
to the halo center regardless of 
the UV strength. The total mass slightly increases from the initial state by accretion of the diffuse gas in the outer part.

\subsection{Mass Loss}  \label{sec:massloss}
The photo-heated gas flows outward from the surface of the self-shielded region, while the central part continues contracting. 
The rate of the gas mass loss can be calculated as 
\eq{
	\dot{M}_{\rm ph} = \int _ S \dd \vec{S} \cdot( \rho_{\rm b} \vec{v} ), 	\label{eq:masslossrate}
}
where 
$S$ is the surface area of the launching base. 
Note that the right-hand-side of this equation is equal to that of \eqnref{eq:masslosseq}.
The mass-loss rate is essentially determined by the radial extent of the self-shielded region, because the initial velocity of the photoevaporative winds is typically the sound speed of the ionized gas ($\sim 10\kms$), and the base density is determined by the EUV flux. The mass flux at the base is not strongly dependent of the gas metallicity \citep{1989_Bertoldi,2019_Nakatani}. 
We measure the total gas mass within a halo as  
\eq{
	M_{\rm b} = \int_{r \leq \rvir} 2\pi R \rho_{\rm b}  \, \dd x \, \dd R.
}
The evolution of $M_{\rm b}$ can be characterized by two phases separated by the time when the diffuse outer part is stripped off and a "naked" dense core is left. In the later phase, the core is directly exposed to UV radiation but 
the net mass loss is small owing to small geometrical size. The   
 photoevaporation rate decreases rapidly during the transition phase. 
 A similar process is known also in the study of molecular cloud photoevaporation \citep[e.g.,][]{1989_Bertoldi, 2019_Nakatani}. 
It is difficult to follow the photoevaporation process
in detail after the transition phase, because
the small core is resolved only with 
several computational cells in our simulations.
We thus calculate $M_{\rm b}$ only up to the transitional phase.

We empirically determine the transition time by the following
conditions:
\gathering{
    \frac{1}{M_{\rm b}}\int_{\rho_{\rm b} > 10^{-3}\rho_{\rm b, max}} 2\pi R \rho_{\rm b}  \, \dd x \, \dd R  
    > 0.8 \label{eq:limitingtime1}\\
    \rho_{\rm b, max} > \rho_{\rm b,0}. \label{eq:limitingtime2}
}
Here, $\rho_{\rm b, max}$ is the maximum density in the computational domain and $\rho_{\rm b,0} \equiv \rho_{\rm b}(t=0,r=0)$.

\begin{figure*}[h]
\begin{center}
\includegraphics[clip, width = \linewidth/2-0.5cm]{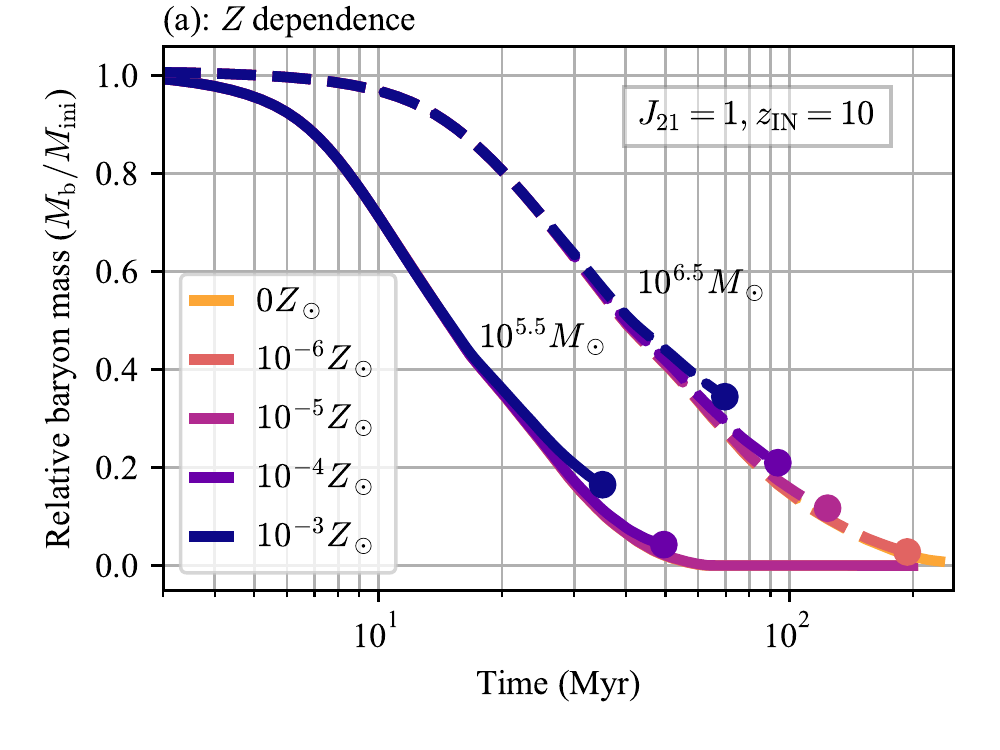}
\includegraphics[clip, width = \linewidth/2-0.5cm]{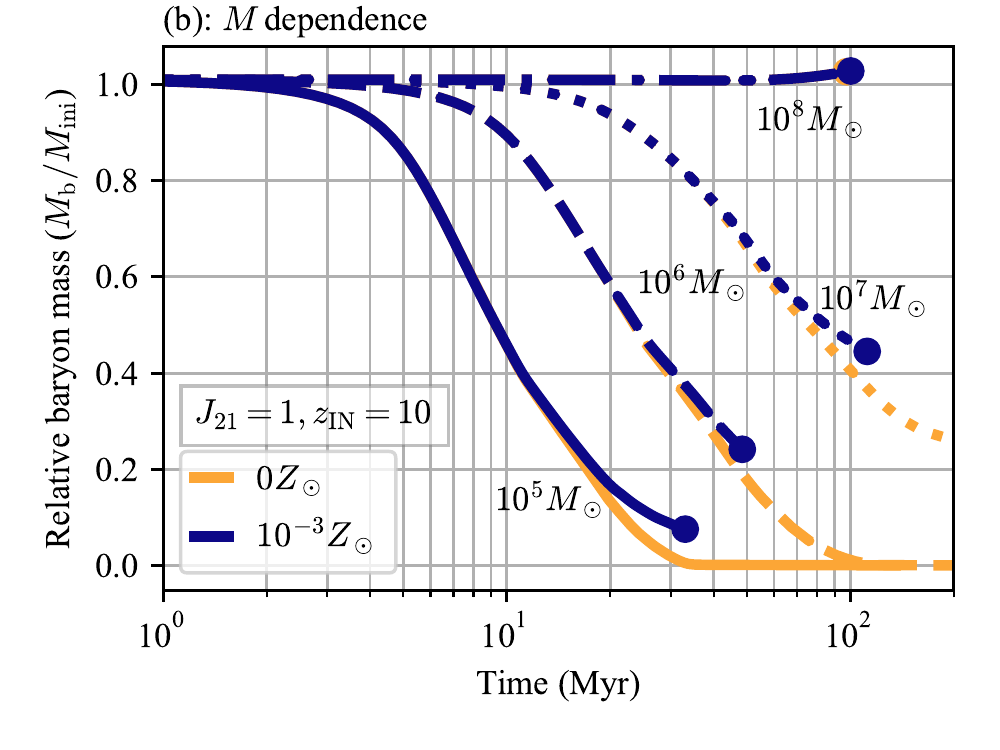}
\includegraphics[clip, width = \linewidth/2-0.5cm]{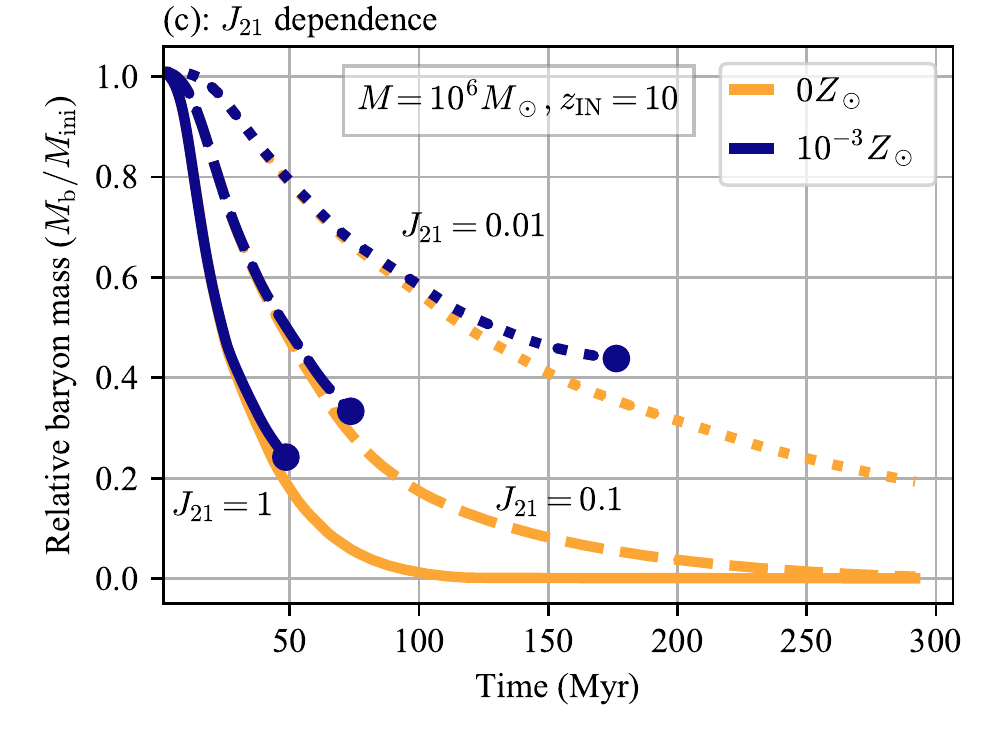}
\includegraphics[clip, width = \linewidth/2-0.5cm]{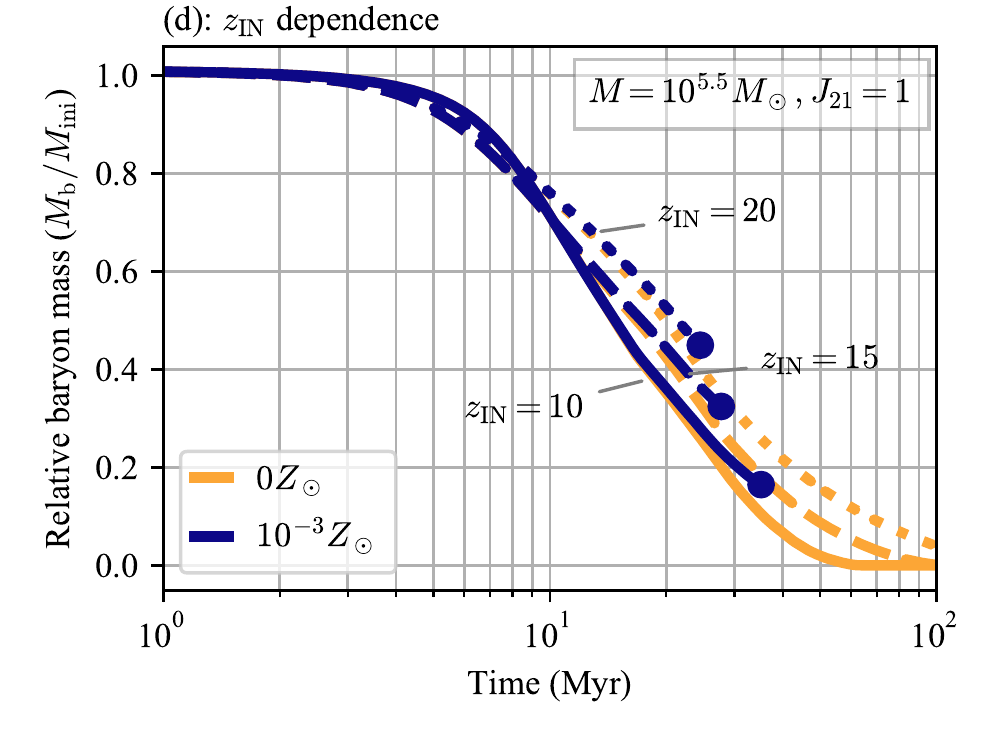}
\caption{
        Time evolution of the total gas mass relative to the initial gas mass for selected runs.
        The panels show the dependence of the gas mass evolution on the four simulation parameters $(\metal, M, J_{21}, \zin)$:
		(a) metallicity dependence of the gas mass evolution for $M = 10^{5.5}, 10^{6.5}$ with $J_{21} = 10 $ and $\zin = 10$. The line colors and styles differentiate metallicity and the halo mass, respectively. The round marker points indicate the time at which all of the atmospheric gas has been lost, leaving an concentrated core (cf.~\eqnref{eq:limitingtime1} and \eqnref{eq:limitingtime2}).
		Note that the curve for \sil{5}{0}{5.5}{10} overlaps those of \sil{$\infty$}{0}{5.5}{10} and \sil{6}{0}{5.5}{10}. 
		(b) halo mass dependence of the gas mass evolution for $\metal = 0, 10^{-3}\,\smetal$ with $J_{21} = 1$ and $\zin = 10$. The line styles correspond to halo masses as annotated. The lines for \sil{$\infty$}{0}{8}{10} and \sil{3}{0}{8}{10} overlap. 
		(c) $J_{21}$ dependence of the gas mass evolution for $\metal = 0, 10^{-2} \, \smetal$ with $M = 10^6 \Msun$ and $\zin = 10$. Solid, dashed, dotted lines indicate $J_{21} = 1, 0.1, 0.01$, respectively. 
		(d) $\zin$ dependence of the gas mass evolution for $\metal = 0, 10^{-3}\,\smetal$ with $M = 10^{5.5}\Msun$ and $J_{21} = 1$. 
		}
\label{fig:masslossrates}
\end{center}
\end{figure*}

\fref{fig:masslossrates} shows the evolution of $M_{\rm b}$ for halos with $M = 10^{5.5-8} \Msun$ with various metallicities and other parameters.
For a given $M$, 
$M_{\rm b}$ evolves along the same track on the $t$--$M_{\rm b}$ plane for various metallicities. 
We have discussed in \secref{sec:result1}
that the effect of metal-enrichment appears clearly in the concentrating core, but
the strength of EUV-driven photoevaporation is independent of metallicity. 
Thus the outer diffuse gas has not cooled in the early evolutionary phase even with nonzero $\metal$ cases, and
the size of the self-shielded region is similar regardless of $\metal$ (compare the rightmost panels in \fref{fig:snapshots}).
Therefore, the evolution of $M_{\rm b}$ does not differ significantly 
until the diffuse envelope gas is lost. 


We show the time evolution of $M_{\rm b}$ for halos with various $M$
including the high mass case ($\Tvir > 10^4\Kelvin$; $M \gtrsim 10^{7.5}\Msun$) 
in \fref{fig:masslossrates} (Panel~b). 
A large amount of gas is lost from low-mass halos ($\Tvir < 10^4\Kelvin$) 
but the mass-loss rates are significantly smaller for massive halos with $10^8\Msun$. 
Interestingly, the gas mass {\it increases} slightly.
The diffuse gas at $r>\rvir$ is accreted while the central part
keeps cooling.
Photoevaporation is hardly observed in these runs, because the initial temperature, $\Tvir$, is higher than the typical temperature of a photo-heated gas in the first place.
Halo's gravity is so strong that it retains the 
photo-heated gas. 
Another important feature is that the gas mass evolution of the massive halos is nearly independent of metallicity. 
Radiative cooling by hydrogen is dominant in these halos (\secref{sec:result1}).

\subsection{Radiation Intensity}	\label{sec:result2}

In photo-ionized regions, the characteristic ionization time,
$t_{\rm ion} \sim 0.01 J_{21}^{-1} \Myr$, is orders of magnitude 
shorter than the typical crossing time of photoevaporative flows,
$ \tcr \simeq \rvir / 10\kms \simeq 10 \,M_{5}^{1/3} [(1+z)/10]^{-1} \Myr$ with $M_{5} \equiv M/10^{5} \Msun$. 
For weak UV radiation, the I-front is located at the outer part 
of the halo where the density is low.
We find that the boundary where $\abn{HII} = 0.5$ is located at 
a radius where $\col{HI} \sim 10^{18} J_{21} ^{1/2} \cm^{-2}$,
and the base density is approximately estimated to be $\nh \sim 10^{-1}$--$10^{-0.5} J_{21} \cm^{-3}$,
which is consistent with the result of \cite{2004_Shapiro}.   

The UV intensity $J_{21}$ does not strongly change the overall dynamical 
photoevaporation process.
However, the mass loss rate sensitively depends on $J_{21}$,
because the base is located at a larger $r$ for lower $J_{21}$, where 
the gas density is lower. 
One may naively expect that the geometrical size can make
the photoevaporation rate large, 
but the small density at the large radius mitigates
increase of mass loss
(cf.~\eqnref{eq:masslossrate}). 
The base density is approximately proportional to the UV flux,
while the base radius increases by only a small factor 
because the density 
rapidly decreases with increasing radial distance.
Hence the mass loss rate, $|\dot{M}_{\rm b}|$, decreases for smaller
$J_{21}$ as can be seen in \fref{fig:masslossrates}-(c). 
The mass loss rate decreases with time, $\dd |\dot{M}_{\rm b}| / \dd t < 0$, because the geometrical cross-section of halo decreases. 
We also note that the characteristic mass-loss time for massive halos 
is longer than the Hubble time 
at the epochs considered. 

\subsection{Turn-on Redshift}	\label{sec:result3}
Halos forming at different redshifts have different properties.
Most notably high-redshift halos are more compact and denser
(cf.~\eqnref{eq:densityprofile}).
We study cases with different 
$\zin$, the timing of radiation turn-on,
when the cosmological I-front reaches the halo.
One can consider that different $\zin$ effectively corresponds 
to different 
reionization histories, or to an inhomogeneous reionization model
in which the effective $\zin$ differs from place to place.

The process of photoevaporation is essentially the same as those described in \secref{sec:massloss} and \secref{sec:result2}.
The gas density at the photoevaporative flow "base" is primarily set by $J_{21}$, and is not explicitly dependent of $\zin$. 
However, the relative distance of the base to the halo center, $\xi (\equiv r/\rvir)$, 
is larger for halos at higher redshift owing to higher average density,
while the size of the neutral, self-shielded region is {\it smaller}.
These two effects nearly cancel out and yield photoevaporation rates nearly independent of $\zin$. 
We find that $|\dot{M}_{\rm b}|$ increases only by $20$--$30\%$ with $\Delta\zin = -5$. 

In the characteristic minihalo case with $M=10^{5.5}\Msun$, $J_{21} = 1$, $\metal = 10^{-3}\,\smetal$ and $\zin = 10$,
about 80\% of the initial gas mass is lost,
and the mass loss fraction is only $\sim 60\%$ with $\zin = 20$. 
Similar trend of the final core mass is seen in other runs with different $M$ and $J_{21}$. This is consistent with the results of
\cite{2005_Iliev} who show weak dependence of mass-loss time scale on turn-on redshift. 

\subsection{Gas Mass Evolution}\label{sec:result4}
We have shown that the gas mass evolution depends most sensitively 
on $M$ and $J_{21}$. Physically, these are the most relevant quantities to the gravitational force and the mass flux, respectively (cf.~\eqnref{eq:masslosseq}).
The results of our numerical simulations can be characterized by two quantities: the half-mass time, $t_{1/2}$, at which the gas fraction decreases to 0.5, and the remaining mass fraction, $f_{\rm b,rem}$, which is the mass fraction of the "remnant" condensed core.
\begin{figure*}[htbp]
    \centering
    \includegraphics[clip, width = \linewidth]{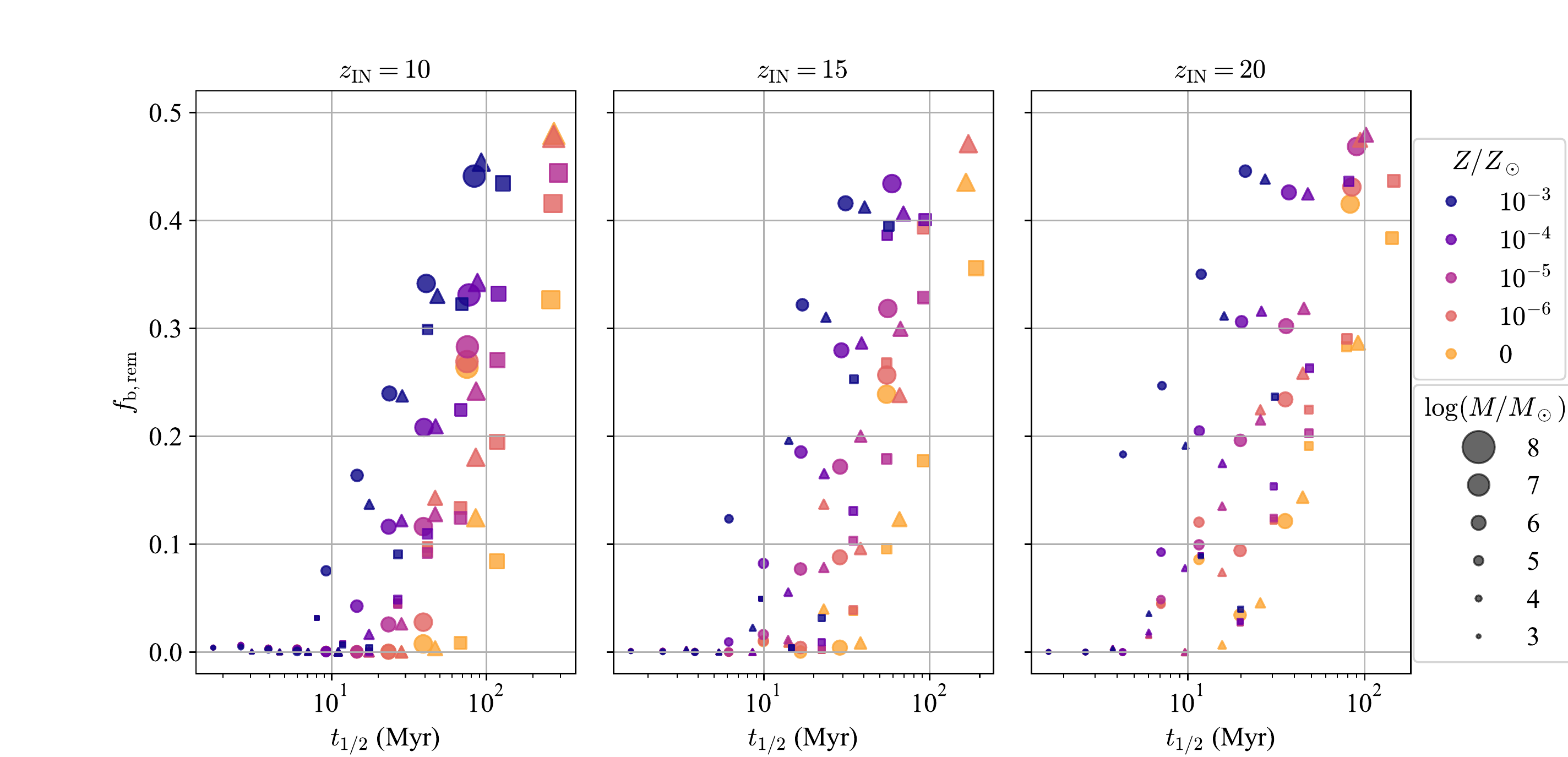}
    \caption{We plot $t_{1/2}$ and $f_{\rm b, rem}$ for our simulated halos. The left, middle, and right panels shows the results at $\zin = 10, 15, 20$, respectively. The circles, triangles, and squares represent $J_{21} = 1, 0.1, 0.01$. The colors of the markers indicate metallicity, and the sizes are scaled according to the halo mass $M$. The size reference is shown at the bottom right. The maximum marker size corresponds to $M = 10^7, 10^{6.5}, 10^{6.5}\Msun$ for the panels of $\zin = 10, 15, 20$, respectively. 
    Note that $t_{1/2}$ is definable only for the halos whose mass reduces to $f_{\rm b, rem} \leq 0.5$. }
    \label{fig:evamap}
\end{figure*}
\fref{fig:evamap} shows the distribution of $t_{1/2}$ and $f_{\rm b, rem}$. 
This summarizes the overall dependence of gas mass evolution on 
the simulation parameters $(\metal, J_{21}, M, \zin)$. 
Massive halos at lower $\zin$ reflect lower average densities.
The mass (symbol size) is larger towards the upper right corner in each panel of \fref{fig:evamap}, indicating that the remaining mass fraction $f_{\rm b, rem}$ increases for higher halo mass. Also halos with higher-metallicity have higher $f_{\rm b, rem}$  owing to
efficient cooling (\secref{sec:massloss}). 
For higher $J_{21}$, both $t_{1/2}$ and $f_{\rm b,rem}$ are smaller
because of the faster mass loss for stronger UV radiation (\secref{sec:result2}).

\subsection{Similarity in Mass Loss}	\label{sec:similarities}
We have shown that the mass-loss rate has weak metallicity dependence
at least until the bulk of the diffuse halo gas is stripped off. In this section, we first derive nontrivial similarity of the gas mass evolution for metal-free halos. We then apply this model to other low-metallicity cases.

In \secref{sec:analytic}, we have developed an analytic model with a key parameter $\chi$ that characterizes the gas mass evolution of a photoevaporating halo. 
There, the effect of the host halo's gravity has not been incorporated (Eqs. \ref{eq:masslosseq}, \ref{eq:nondimmassloss}, and \ref{eq:difmassloss}). 
We expect that deceleration by gravity becomes important for halos
whose virial temperatures are 
comparable to the typical temperature of the photo-ionized gas, $\sim 10^4\Kelvin$. In such cases, assuming $c_i = 10\kms$ as the photoevaporative
flow velocity overestimates the photoevaporation rate (see \eqnref{eq:difmassloss} and the description above it). 
To account for the deceleration, we adopt
a "reduced" parameter defined as 
\eq{
    \chi^\prime \equiv \chi \frac{c_i - V_{\rm c}}{c_i},    \label{eq:modchi}
}
in the following discussions. The derived values are listed in \tref{tab:data} of \appref{sec:supplymental}. 

\begin{figure}[htbp]
\begin{center}
\includegraphics[clip, width = \linewidth]{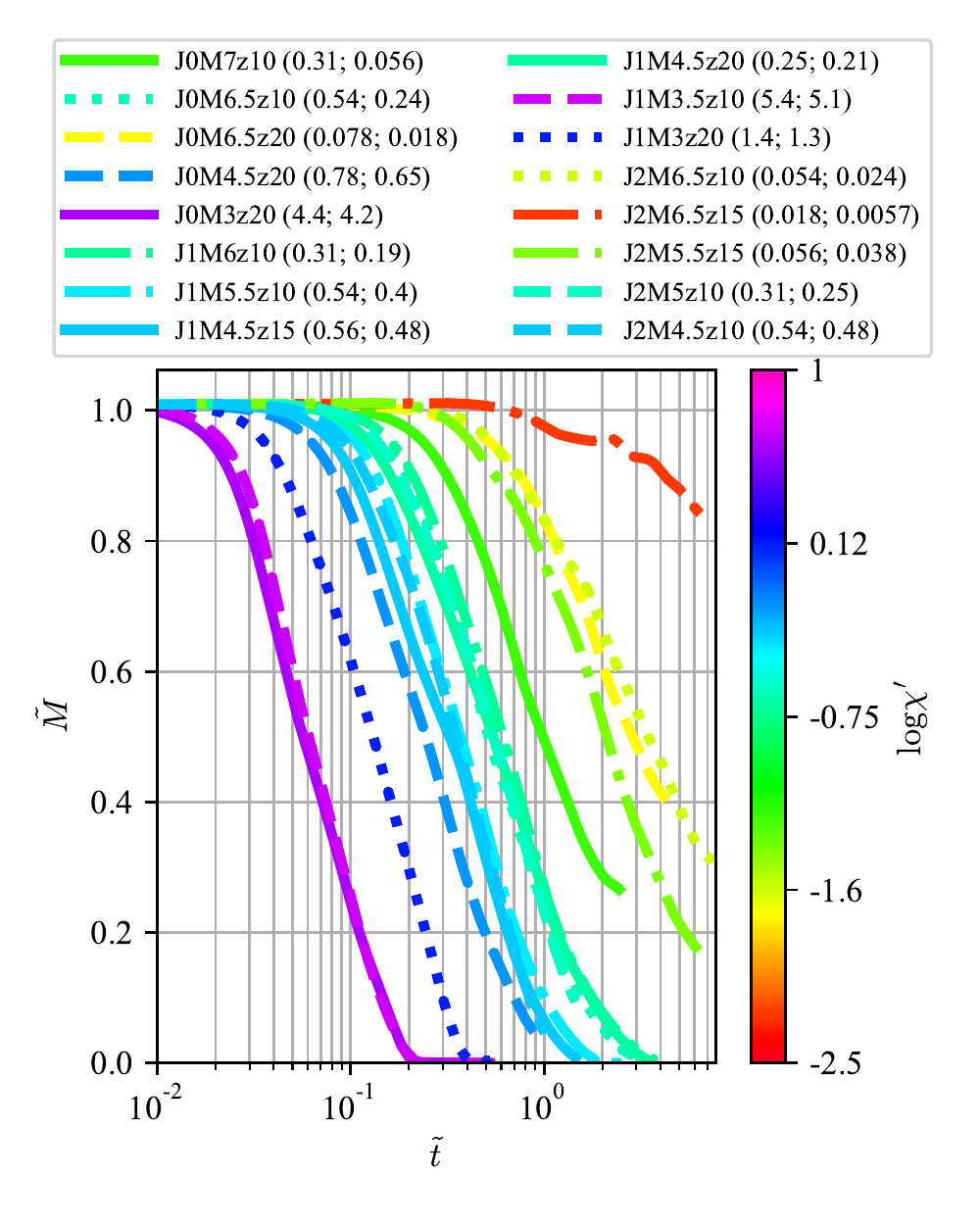}
\includegraphics[clip, width = \linewidth]{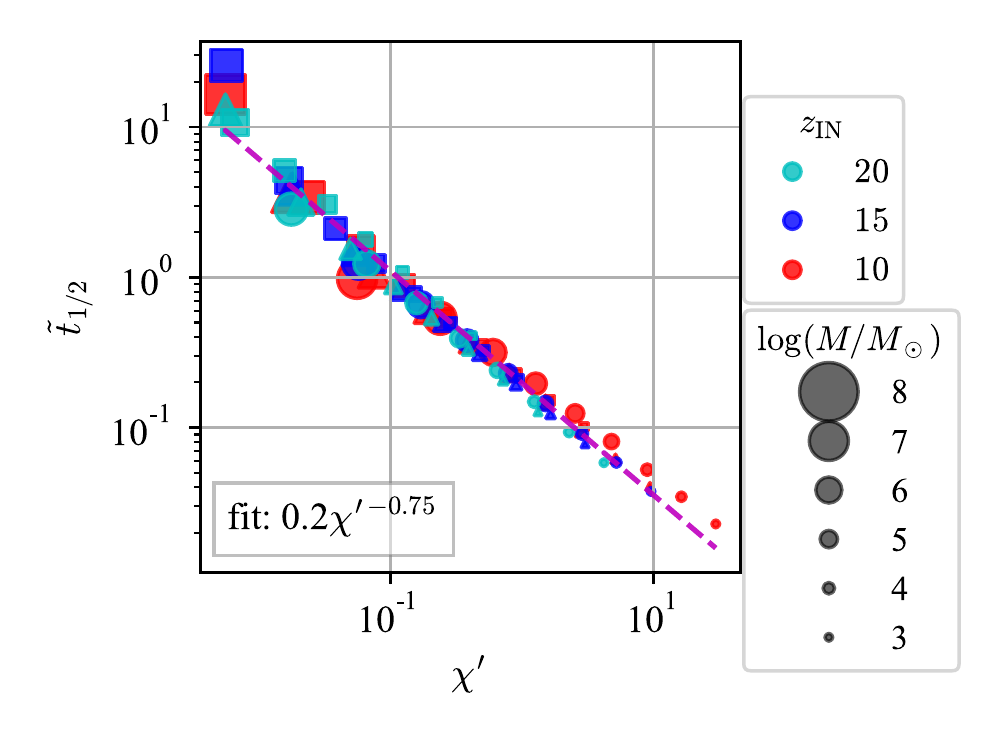}
\caption{
(top) Similarities in the gas mass evolution for metal-free halos with various parameter sets. Note that we omit ``Z$\infty$'' from the simulation labels in the legend. Values in the parentheses after each simulation label show corresponding $\chi(\equiv \eta q^{-0.5})$ and $\chi^\prime (\equiv \eta q^{-0.5} \Delta c_i / c_i)$. The horizontal and vertical axes indicate the dimensionless time and mass, respectively (cf.~\eqnref{eq:nondimmassloss}). Lines are colored according to $\chi^\prime$ as indicated by the color bar. 
Lines with similar colors almost overlap on the $\tilde{t}$--$\tilde{M}$ plane, indicating that having similar $\chi^\prime$ results in similar mass evolution. 
(bottom) $\chi^\prime$ vs dimensionless half-mass time $\tilde{t}_{1/2}$ for metal-free halos. The markers are used in the same manner as in \fref{fig:evamap} but the colors represent $\zin$. The magenta dashed line is a fit. A negative correlation is seen between $\chi^\prime$ and $\tilde{t}_{1/2}$. 
}
\label{fig:similarity}
\end{center}
\end{figure}

The top panel of \fref{fig:similarity} shows the gas mass evolution of metal-free halos with various parameter sets in the dimensionless form. 
Halos with similar $\chi$ or $\chi^\prime$ 
evolve on essentially the same track in the $\tilde{M}$--$\tilde{t}$ plane.
We explain the similarity further by using a specific example as follows.
The simulation parameters of \sil{$\infty$}{0}{6.5}{10} (cyan dotted line; $\chi = 0.54, \chi^\prime = 0.24$) are close to those of \sil{$\infty$}{0}{6.5}{20} (yellow dashed line; $\chi = 0.078, \chi^\prime = 0.018$), but the mass evolution significantly deviates from each other on the dimensionless plane. 
On the other hand, it is closer to those of \sil{$\infty$}{1}{4.5}{20} (cyan-green solid line; $\chi = 0.25, \chi^\prime = 0.21$) and \sil{$\infty$}{2}{5}{10} (cyan dashed line; $\chi = 0.31, \chi^\prime = 0.25$). 
A more straightforward case is \sil{$\infty$}{2}{5}{10} (cyan solid line; $\chi = 0.31, \chi^\prime = 0.25$). It is close to \sil{$\infty$}{2}{4.5}{10} (light-blue dashed line; $\chi = 0.54, \chi^\prime = 0.48$), as expected from the close parameter values. Interestingly, the \sil{$\infty$}{2}{4.5}{10} (blue dashed line; $\chi = 0.54, \chi^\prime = 0.48$) run is also very close to  \sil{$\infty$}{1}{5.5}{10} (cyan dash-dotted line; $\chi = 0.54, \chi^\prime = 0.4$).

We find that gravitational deceleration of the photoevaporating gas is 
an important factor.
The evolution in Run~Z{$\infty$}J{0}M{6.5}z{10} (cyan dotted line) is close to 
other cases with green lines, but its $\chi$ value $(\approx 0.54)$ is actually closer to those of runs indicated by the blue lines. 
Also, $\chi \approx 0.31$ for Run \sil{$\infty$}{0}{7}{10} (yellow-green solid line)
is close to those of runs indicated by the green lines, but the 
actual evolution apparently deviates. 
Gravitational deceleration of the photoevaporating gas reduces the mass-loss rate 
for these relatively massive minihalos with virial temperature of $2400 - 5000 \Kelvin$. 
Clearly, it is important to incorporate the correction of $\chi$ owing to deceleration.
We conclude that $\chi^\prime$ is the essential parameter to characterize the gas mass 
evolution of photoevaporating halos.

The bottom panel of \fref{fig:similarity} shows correlation between $\chi^\prime$ and the dimensionless half-mass time $\tilde{t}_{1/2} \equiv t_{1/2} / t_0$ for metal-free halos. There is a tight correlation
given by $\tilde{t}_{1/2} = 0.2 {\chi^\prime}^{-0.75}$. The correlation confirms the importance of $\chi^\prime$ in characterizing the gas mass evolution of photoevaporating halos. 
Note that the same correlation holds for low-metallicity halos.
Thus the fit can be applied for halos with any metallicity that lose more than a half of the initial mass. 

\subsection{Fitting function}	\label{sec:evatime}
Based on the similarity studied so far,
we derive a fit of $\tilde{M} $ that can be readily used 
in semi-numerical models \citep[e.g.,][]{2013_Sobacchi, 2012_Fialkov, 2013_Fialkov, 2015_Fialkov, 2016_Cohen}. 
From the result shown in ~\fref{fig:masslossrates},
we propose a function
\eq{
\splitting{
	\tilde{M}_{\rm fit} &= f (\tilde{t}) =  \frac{1-C_1}{(\tilde{t}/\tilde{t}_{\rm s})^{p} + 1} + C_1, \label{eq:fittingfunc}
	}
}
where $p, C_1$, $\tilde{t}_{\rm s}$ are fitting parameters,
which control the steepness of mass decrease, the remaining mass fraction, and the dimensionless time at which $f_{\rm b} = (1+C_1)/2$, respectively. 
We restrict the parameter ranges to $0 \leq C_1\leq 1$, $0 \leq \tilde{t}_{\rm s}$, and $0\leq p$, in order to avoid unphysical fitting results. 

\begin{figure}[htbp]
\begin{center}
\includegraphics[clip, width = \linewidth]{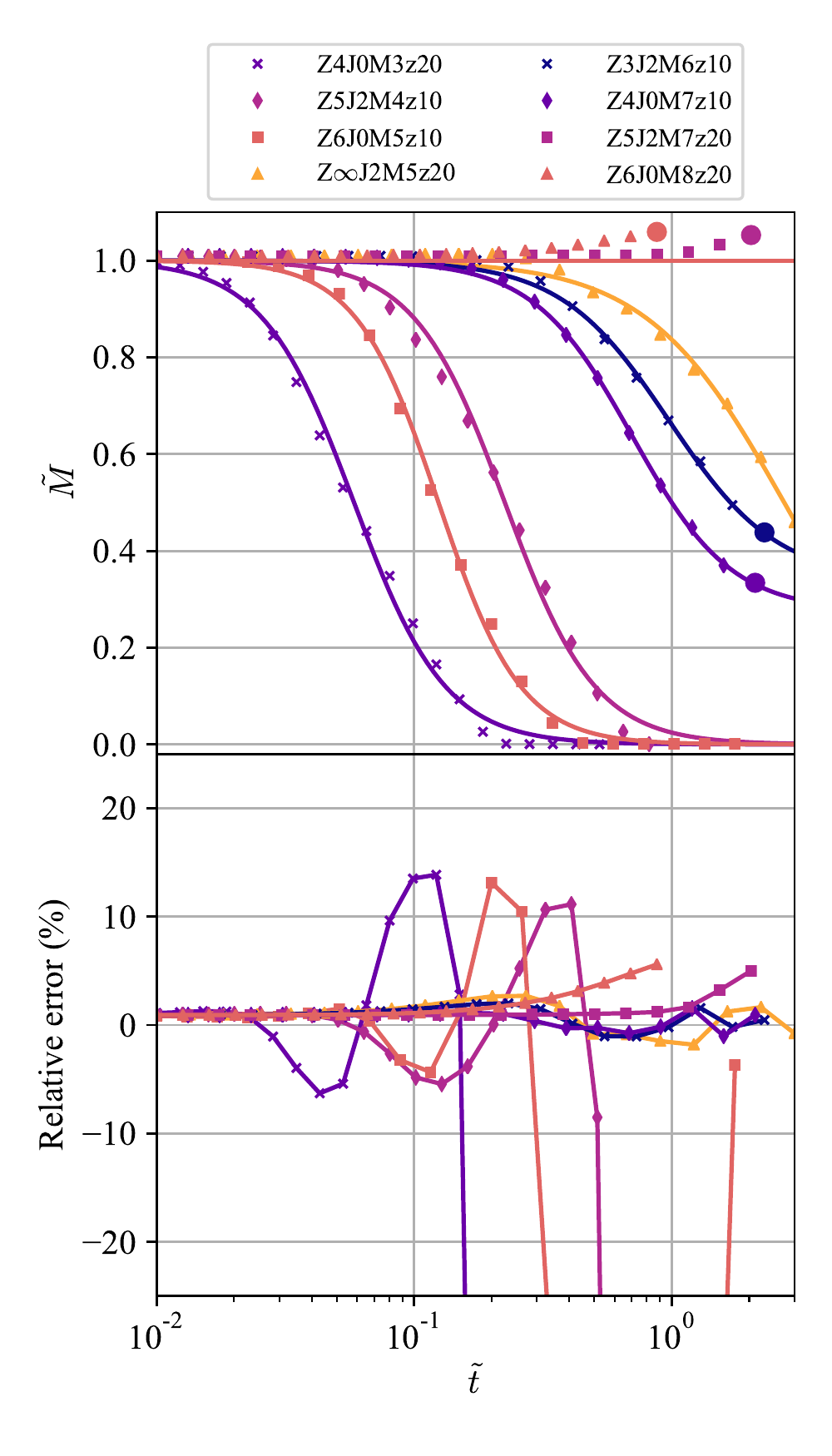}
\caption{
Top: We compare the mass evolution given by equation (\ref{eq:fittingfunc}) and the 
simulation results. 
The horizontal and vertical axes are the same as \fref{fig:similarity}. The solid lines show fits, and the marker points show the mass evolution in several selected runs. The marker's color represents metallicities, but the marker shapes are randomly adopted. 
The round marker points at the tails of the curves for Z3J2M6z10, Z4J0M7z10, Z5J2M7z20, and Z6J0M8z20 indicate the time at which the bulk of the atmospheric gas is lost as in \fref{fig:masslossrates}.
Bottom: Relative errors between the simulation and the fit in percent. 
}
\label{fig:fitting}
\end{center}
\end{figure}

We list the best fit values in \tref{tab:data} in Appendix.
The excellent accuracy of the fit can be seen in \fref{fig:fitting}
in comparison with the simulation results.
The three-parameter function fits the simulation results well until the mass decreases to $\tilde{M} \approx 0.01$. 
For halos with $\Tvir > 10^4\Kelvin$,
the gas mass increases owing to the halo's strong gravity. 
We do not consider the slight increase of gas mass in deriving the fit of \eqnref{eq:fittingfunc},
and simply set $\tilde{M}_{\rm fit} = 1$.

Since we have not followed the evolution of the dense core after the 
outer diffuse gas is photoevaporated, the resulting 
$\tilde{M}_{\rm fit}$ does not strongly depend on metallicity.
The photoevaporation becomes inefficient in the late phase when the concentrated core is directly exposed to the UVB radiation, because of its small geometrical cross section (cf.~\eqnref{eq:masslosseq}).
In order to follow the late phase evolution more accurately, 
we will need to run simulations with
a much higher spatial resolution so that the small core can be fully resolved.
However, we expect that the mass evolution would not differ significantly 
from those obtained in this section, 
because the remaining mass fraction is already small in the late phase. 

\section{Discussion} 	\label{sec:discussion}

\subsection{Star Formation in Metal-Enriched Halos}
\label{sec:starformation}

	As the surrounding gas cools and falls on to the center, the
	central gas density further increases but the 
    temperature remains low, 
    and thus the dense core can be gravitationally unstable to induce 
    star formation.
	In metal-enriched halos, cooling by metal atoms/ions and by dust grains
	enable the gas to condense even under the influence of strong UV radiation.
	Hence, metal enrichment can effectively lower the mass threshold of star-forming halos \citep[$\sim 10^5$--$10^6\Msun$; e.g.,][]{1997_Tegmark, 2001_Machacek, 2002_BrommCoppiLarson, 2003_YoshidaAbelHernquist}. 
    In this section, we study the relation between metallicity and the minimum star-forming halo mass, $M_{\rm min}$. 
	
\begin{figure*}
	    \centering
	    \includegraphics[clip, width = \linewidth]{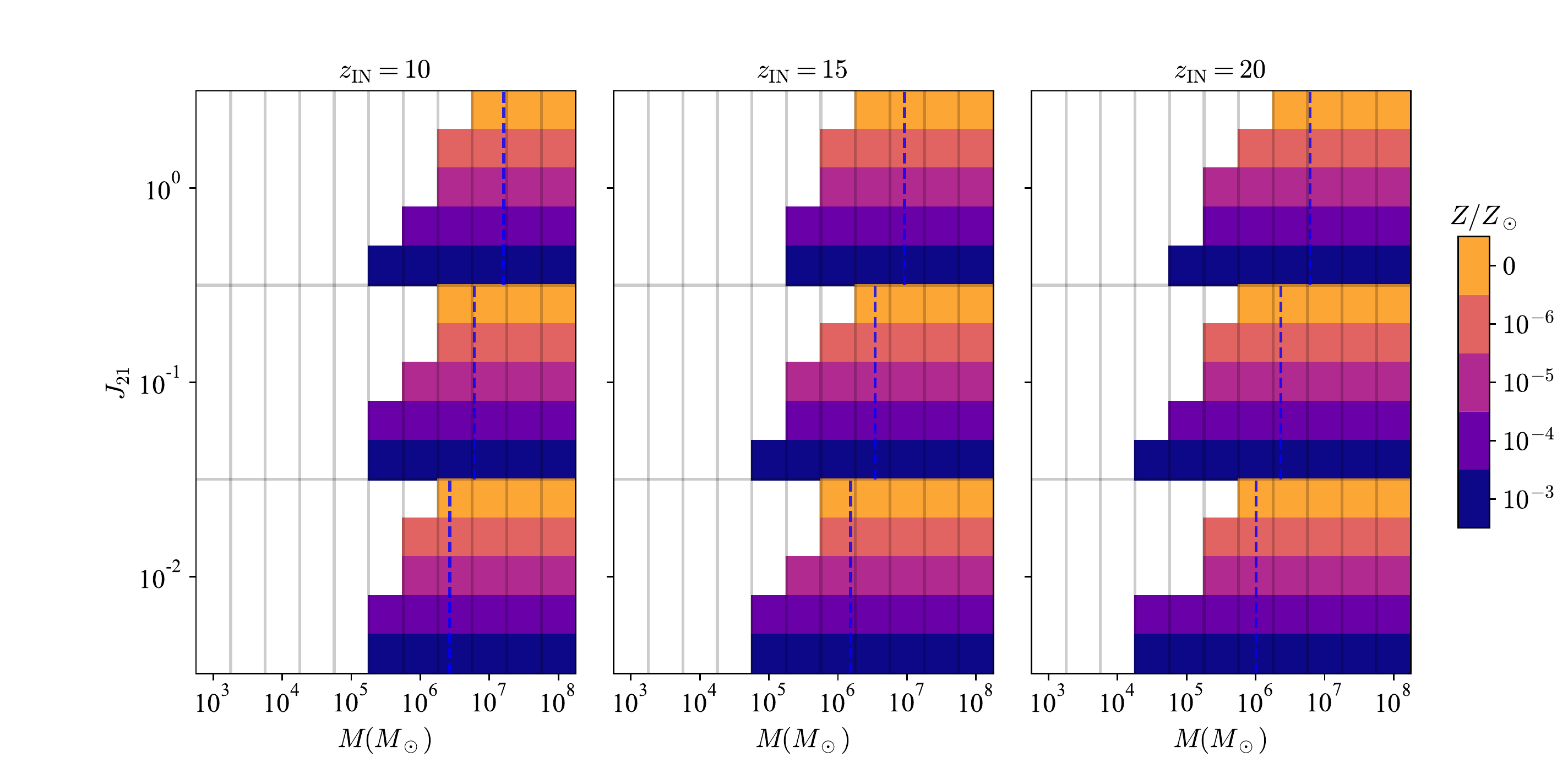}
	    \caption{Star formation vs. photoevaporation for all our runs.
	    The left, middle, and right panels correspond to $\zin = 10, 15, 20$, respectively. 
	    The horizontal and vertical axes are halo mass and $J_{21}$ in all the panels. 
	    Each of the three panels is divided into $11\times 3$ rectangular blocks indicating combinations of $M$ and $J_{21}$. Each block is further divided into 5~square sections to show the dimension of metallicity; corresponding metallicity is $\metal = 0,10^{-6},10^{-5},10^{-4},10^{-3}\,\smetal$ from top to bottom.
        We represent star-forming halos by filling the sections with colored squares as
	    $10^{-3}\,\smetal$ (navy), $10^{-4}\,\smetal$ (purple), $10^{-5}\,\smetal$ (pink), $10^{-6}\,\smetal$ (orange), $0\,\smetal$ (yellow). 
	   The vertical blue dashed line shows the molecular cooling limit, which we derive using an expression in \cite{2013_Fialkov} \citep[cf.][]{2001_Machacek, 2012_Fialkov}, for a reference; note that we have not taken into account the correction to the molecular cooling limit due to relative velocity between baryons and cold dark matter. Incorporating the correction does not significantly change the limit mass for the redshifts of interest here.
	    }
	    \label{fig:starformation}
	\end{figure*}

	We assume that star formation occurs when an enclosed gas mass
	\eq{
		M_{\rm enc} ( r) = \int_{\leq r} \rho_{\rm b} \,  \dd V .
		\label{eq:enclosedmass}
	}
	exceeds the Bonnor-Ebert mass \citep{1955_Ebert, 1956_Bonnor},
	\eq{
		M_{\rm BE}  ( r) \simeq 1.18 \frac{\bar{\cs}^4}{\sqrt{P_{\rm c} G^3}},  \label{eq:bemass}
	}
	at a spherical radius of $r$.
	Here, $\bar{\cs}$ is an average sound speed of gas,
	and $P_{\rm c}$ is confining pressure. 
	We calculate $\bar{\cs}$ and $P_{\rm c}$ 
	by integrating the pressure within an enclosed volume $V$ and over the corresponding enclosing surface $\partial V$, respectively,
	\gathering{
		\bar{\cs}  (r) =   \sqrt{ M_{\rm enc} ^{-1} \int_{\leq r }P \, \dd V } \\ 
		P_{\rm c}  (r)  =  \dfrac{1 }{4 \pi r^2} \int_{\partial V}  P \, \dd S.
		\label{eq:confiningpressure}
	}
	Since we are interested in star formation within dark halos,
	we set the enclosing radius to the scale radius (core radius), $r_{\rm s} \equiv \rvir/c_{\rm N}$, in \eqnref{eq:enclosedmass}--\eqnref{eq:confiningpressure}. 
	We regard a halo as star-forming if it has $M_{\rm enc}(r_{\rm s})/M_{\rm BE}(r_{\rm s})$ larger than unity at a certain point during the evolution. 
	\fref{fig:starformation} shows star-forming halos and non-star-forming halos defined by this condition. 
	We also provide a fit to the resulting $M_{\rm min}$ as a function of $\zin$, $J_{21}$ and $\metal$ in \appref{sec:fitminimum}. 
	{}
	
	In molecular cooling halos, the gas cools to satisfy the unstable condition, $M_{\rm enc}/M_{\rm BE}>1$, at any metallicity including the primordial case. 
	The halos retain more than 10\% of the initial gas. 
	The remaining gas mass is larger for higher halo mass and for lower $J_{21}$. 
	It is nearly unity for atomic cooling halos ($\Tvir > 10^4\Kelvin$). 
	
	We find strong impact of metal enrichment in halos whose mass is lower than the atomic cooling limit. 
    The gas cools via \ce{H2} cooling faster than the bulk of the gas is photoevaporated. 
	This effect is clearly seen in higher-metallicity halos (\fref{fig:starformation}).
	Interestingly, the minimum collapse mass is lowered 
	even with very small metallicities ($\metal \lesssim 10^{-5}\,\smetal$),
	and becomes as small as $M_{\rm min} \sim 10^4\Msun$ with $\metal \gtrsim 10^{-4}\,\smetal$. 
	
    \citet{2014_Wise} show that 
    star formation is active in the metal-cooling halos, and the ionizing photons of the formed stars can provide up to 30\% of ionizing photons responsible for reionization. 
    Metal-cooling halos are {\it heavy} analog of molecular cooling halos and have masses slightly below the atomic cooling limit ($\sim 10^8\Msun$). We have found {\it light} analog of molecular cooling halos in which the gas cools by \ce{H2} molecules that are formed by grain-catalysed reactions. 
    Recent numerical simulations suggest formation of massive or even very massive stars in metal-enriched halos \citep{2016_Chiaki, 2018_Fukushima}. 
    Theses stars can allow the beginning of reionization to occur earlier than without the metal effects. It does not likely change the redshift where reionization completes \citep{2018_Norman}.

\subsection{Implications to $21\cm$ Line Observations}
The hyperfine spin-flip emission of atomic hydrogen (the so-called $21\cm$ line) is a promising probe of the neutral IGM in the Epoch of Reionization. 
Because the strength of the 21-cm signals depend crucially on astrophysical processes at the early epochs, observations of the emission/absorption against the CMB will provide invaluable information on star formation and the physical state of the IGM. 

 Semi-numerical models have been used to predict the large-scale fluctuations of  the $21\cm$ signal \citep[e.g.,][]{Mesinger:2011,Fialkov:2014b}.
  Results of such models are often utilized to derive upper limits on the high-redshift astrophysics \citep[e.g.,][]{Monsalve:2018, Monsalve:2019,Ghara:2020,Mondal:2020, Greig:2020}, and make forecasts for ongoing  measurements of the 21-cm power spectrum with  experiments such as HERA \citep{DeBoer:2017}, LOFAR \citep[e.g.,][]{Mertens:2020}, MWA \citep[e.g.,][]{Trott:2020}, LWA \citep{Eastwood:2019}, and the future SKA \citep{Koopmans:2015}, and of the 21-cm sky-averaged (global) signal using LEDA \citep{Price:2018}, SARAS \citep{Singh:2018}, EDGES \citep{Bowman:2018}, PRIZM \citep{philip19}, MIST\footnote{http://www.physics.mcgill.ca/mist/} , and REACH \footnote{https://www.kicc.cam.ac.uk/projects/reach}. 
  
  The speed and convenience of the semi-numerical methods come along with their poor spacial resolution which is compensated by extensive use of sub-grid models. 
For example, \citet{2013_Sobacchi, 2016_Cohen} study the effect of the UV
background radiation in terms of gas cooling threshold $M_{\rm cool}$. 
Halos more massive than $M_{\rm cool}$ are regarded as star-forming halos. 
The simple prescription in previous studies, however, does not take into account the effect of metal-enrichment. 
As we have shown in \secref{sec:starformation}, star formation can occur in metal-enriched minihalos even during reionization.
\cite{2016_Cohen} have shown that star formation in such metal-enriched minihalos affects the $21\cm$ signal from high redshifts.
In particular, signatures of baryon acoustic oscillation (BAO) imprinted on the $21\cm$ signal is amplified and can possibly be detected over a wide range of redshift. 
\cite{2016_Cohen} set a threshold mass for cooling (and thus star-formation) similar to that of molecular cooling halos. 
Interestingly, this somewhat conservative assumption is well justified by our results, where metal enrichment lowers $M_{\rm min}$ from the molecular cooling limit by an order of magnitude for $\metal \gtrsim 10^{-4}\,\smetal$ even under the effects of UVB. Our results support the predicted enhancement and survival of BAO signature imprinted on $21\cm$ signals due to effects of metal enrichment. 

\subsection{Internal Stellar Feedback Effect}

	Massive stars formed in metal-enriched halos affect the host halo by UV radiation and stellar winds, and by supernova explosions.
	Then the halo gas would not only photoevaporate by external UV radiation
	but also can be dispersed by these internal processes.
	
	The stellar feedback is effective if
	(i) the enclosed gas mass $M_{\rm enc}$ (\eqnref{eq:enclosedmass})
	exceeds the Bonnor-Ebert mass $M_{\rm BE}$ (\eqnref{eq:bemass});
	(ii) stellar feedback energy deposited by massive stars, $E_{\rm dep}$,
	is larger than the gravitational binding energy of the neutral gas
	in the self-shielded regions
	\eq{
		E_{\rm dep} \gtrsim \frac{G (M + M_{\rm s})  }{r_{\rm s} } M _{\rm s},
	}
	where $r_{\rm n}$ is the size of the neutral gas clump.
	We first consider only the supernova explosion energy for simplicity. 
	The deposited energy by other feedback processes can be easily accounted for
	by increasing $E_{\rm dep}$ by a suitable factor.
	
	The cold gas that satisfies the condition~(i) is assumed to form stars with 
	a star formation efficiency of $c_*$
	and with an initial mass function, $\Psi(M_*)$.
	Let $\epsilon_{\rm SN}$ be the average 
	supernova explosion energy. 
	The deposited feedback energy is estimated as 
	\eq{
		E_{\rm dep} = c_* M_{\rm cold} \, \epsilon_{\rm SN}
		\braket{\int M_* \Psi \dd M_*}^{-1}
		{\int_{ M_{\rm th} }\Psi \dd M_*},
	}
	where $M_{\rm cold}$ is the mass of enclosed cold gas,
	and $M_{\rm  th}$ is a threshold stellar mass above which stars cause supernova explosion.  
	We adopt $c_* = 0.1$, $\epsilon_{\rm SN} = 10^{51} \erg$,
	$M_{\rm th} = 8\Msun$, 
	and use the initial mass function of \cite{2003_Chabrier}.

\begin{figure}[h]
\begin{center}
\includegraphics[clip, width = \linewidth]{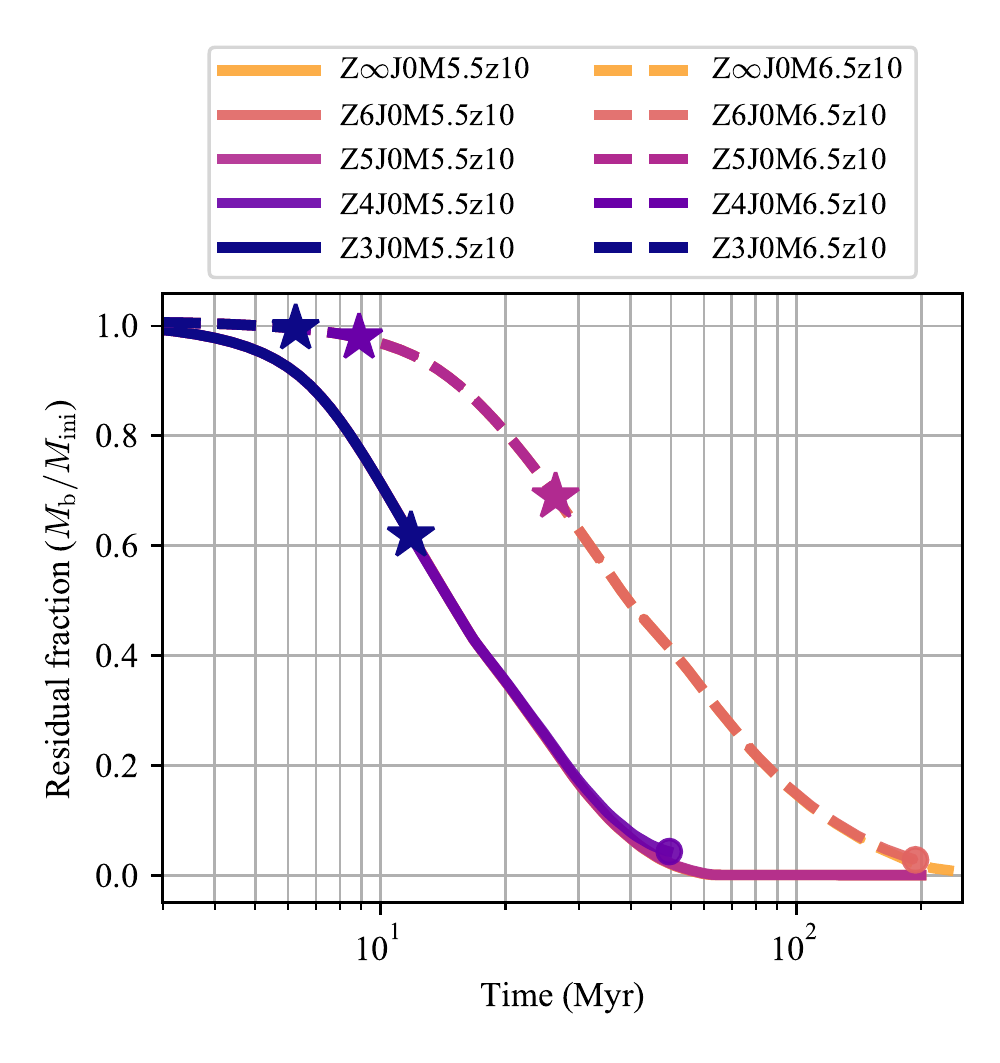}
\caption{
		The same plot as \fref{fig:masslossrates}-(a), 
		but the lines are terminated at the time of star formation and feedback (marked with star symbols) 
		for halos that meet the conditions~(i) and (ii). 
				}
\label{fig:masslossZdependence_stars}
\end{center}
\end{figure}

When the conditions~(i) and (ii) are met and stars are formed 
over a free-fall time, 
the self-shielded region is expected to evolve differently 
from what has been shown in \secref{sec:results}.
If the feedback effects are strong enough to disrupt the entire halo, 
the halo evolution described in \secref{sec:results} 
is valid only up to one free-fall time. 
\fref{fig:masslossZdependence_stars} shows the same plots as \fref{fig:masslossrates}-(a), 
but the lines extend to the time when the stellar feedback is assumed to be effective.
Massive, metal-rich halos are likely to be self-destructed before the gas is lost via photoevaporation. 
Note that the fraction of the cold gas that is converted to stars is small.

In metal-free halos, the stellar feedback can also be important, especially for massive halos. 
We expect that the stellar feedback is unimportant
in low-mass halos with $\Tvir \lesssim 100\Kelvin$ ($M\lesssim 10^{4.5} \,\Msun$ at $z = 10$) for any metallicity. In these halos, most of the gas is quickly lost by photoevaporation before star formation.
In conclusion, 
the deposited energy due to the stellar feedback can disperse the gas from the host halo,
similarly to the well-known feedback effect in dwarf galaxies devoid of \ion{H}{1} content 
\citep{2009_Grcevich, 2014_Spekkens}.

\subsection{X-Ray Effects}
X-rays are attenuated by a larger column $(\sim 10^{21} \cm^{-2})$ compared to EUV \citep{2000_Wilms}.
They can reach halos and pre-ionize/pre-heat the gas 
before the ionization front hits the halos. 
A larger attenuation column also implies larger penetration depths. 
Higher-density photoevaporative flows would be driven 
if the gas temperature increases sufficiently to allow it escape from gravitational binding. 
Accordingly, mass-loss rates would be significantly larger than those of EUV-driven photoevaporation. 

X-ray possibly delays concentration of the self-shielded regions and thereby star formation, if it efficiently heats the gas. 
On the other hand, X-ray ionization can promote \ce{H2} formation via the electron-catalyzed reactions: 
\ce{H + e- -> H- + \gamma} and \ce{H- + H -> H2 + e-} \citep{1996_Haiman, 1999_Bromm, 2003_Glover, 2015_Hummel,2011_Inayoshi,2015_InayoshiTanaka, 2016_Glover, 2016_ReganJohanssonWiseb}. 
If X-rays are strong, very strong LW intensities are required to photodissociate \ce{H2} in the entire halos. 
We expect X-rays to have significant effects on evolution of irradiated halos and star formation activities. 
X-ray chemistry is already implemented in our code \citep{2018_Nakatanib} and we plan to investigate their influences on halo photoevaporation in the future.

\subsection{Model Limitation}

We have studied the gas mass evolution for a wide range of model parameters. 
Exploring the large volume of the parameter space is essential in order to 
understand the evolution of a {\it population} of photoevaporating halos during reionization. Since we have adopted a few simplifications mainly to save computational times, it would be worth examining the limitations of our results. 
We have fixed the dark matter halo potential throughout our simulations. 
In practice, halos grow in mass by mergers and accretion, which in turn strengthens the halo's gravity \citep[e.g.,][]{2013_SobacchiMesinger}. 
This effect may not be negligible for the halos that have longer mass-loss timescale than halo's growth timescale.
Especially, our results indicate that photoevaporation will be significantly suppressed, once halos grow to $T_{\rm vir} \gtrsim 10^4\Kelvin$. 
Lower mass halos ($T \ll 10^4\Kelvin$) disperse quickly, and thus including halo growth is not expected to significantly alter our results. In fact, we find that our results at $\metal = 0\,\smetal$ 
are in good agreement with \cite{2004_Shapiro} and \cite{2005_Iliev}, 
where halo evolution is incorporated in an approximate manner. 

The UVB radiation intensity is also fixed in our simulations.
In general, it can vary over time.
Although the UV background intensity is hardly constrained by observations
for $z\gtrsim 5$ \citep[e.g.,][]{2007_BoltonHaehnelt, 2011_McQuinn}, numerical simulations predict that the UV background builds up over 
several hundred million years. As we have shown, $J_{21}$ strongly affects the mass evolution of halos at any metallicity (\fref{fig:masslossrates}-(c)). 
We expect that growth of halo with time partially compensates for the rise in the intensity of the ionizing background; 
we have shown that mass of halos giving similar $\chi^\prime$, which is approximately proportional to $J_{21}^{1/2} M^{-1/2} (1+z)^{-3}$, evolve in a similar manner (cf.~\secref{sec:similarities} and Eqs.(\ref{eq:chi}), (\ref{eq:modchi})). 

Finally, metallicity is also fixed in each individual run of our simulations. Cosmic average metallicity can increase by an order of magnitude on a timescale of $\sim 1{\rm \,Gyr}$ \citep{2014_MadauDickinson}, 
which is a much longer that typical mass-loss timescales that we found in this study.  
Time-dependent metallicity might have an effect only on long-lived photoevaporating halos by affecting their thermochemistry.
However, we expect that the metallicity-dependent trend would not significantly differ from what we have reported in this study.

\section{Conclusions and Summary}	\label{sec:conclusions}
Photoheating and metal enrichment of halos during the epoch of reionization can affect star formation efficiency and thus change the course of reionization. The effects of metal enrichment, however, have never been explored systematically  in prior works. We have run a suite of hydrodynamics simulations of photoevaporating minihalos ($\Tvir\lesssim10^4\Kelvin$) irradiated by the UVB, covering a wide range of metallicity, halo mass, UV intensity, and turn-on redshift of UV sources. 

Our main findings are summarized as follows: 
\begin{itemize}
    \item 
    In low-mass minihalos with $\Tvir < 10^4\Kelvin$,
    the gas cools mainly via \ion{C}{2} and \ce{H2} line emission to a temperature of $\lesssim 100\K$ if $\metal \gtrsim 10^{-4}\,\smetal$.
    \ce{H2} molecules are produced through grain-catalyzed reactions in 
    the self-shielded, neutral region.
    The cooled gas concentrates towards the potential center to form a dense core. 
    
    \item The evolution of the gas mass is qualitatively the same at any metallicity. The photoevaporation rate decreases after the bulk of the 
    diffuse gas is lost. The dispersal of the diffuse gas completes at an earlier time for halos with higher metallicity.

    \item  In halos with $\Tvir > 20000\Kelvin$, the gas cools by hydrogen Lyman$\alpha$ cooling, and thus the overall evolution of photoevaporating halos does not depend on metallicity. 

    \item The photoevaporation rate depends only weakly on turn-on redshift, and it is slightly smaller for higher $z_{\rm IN}$.

    \item There is a simple scaling relation for the gas mass evolution of photoevaporating minihalos. 
    The time evolution is characterized by a parameter ($\chi^\prime$) scaling as $\propto J_{21}^{1/2} M^{-1/2} (1+z)^{-3}$. 
    It indicates that the obtained evolution applies to any halos with the same $\chi^\prime$. 
    We give a fit to the gas mass evolution as a function of time.   
    
    \item   The concentrating cores of the molecular/atomic cooling halos are likely a suitable environment for star formation. The efficient cooling of metal-rich halos fastens the concentration, and it results in lowering the molecular cooling limit by a factor for small metallicities ($\metal \lesssim 10^{-5}\,\smetal$) and by an order of magnitude for metal-rich cases ($\metal \gtrsim 10^{-4}\,\smetal$). 
    Stellar feedback of the formed stars may be significant enough to disperse the baryons of the parental molecular cooling halos.  

\end{itemize}

Our study suggests existence of small mass, metal-enriched halos
in which stars are formed even under the influence of emerging
UV background radiation.

\acknowledgments %
We thank Gen Chiaki and Kana Moriwaki for insightful comments 
on this manuscript and technical advice. 
RN acknowledges support from Special Postdoctoral Researcher program at RIKEN and from Grant-in-Aid for Research Activity Start-up (19K23469).
AF is supported by the Royal Society University Research Fellowship.
The numerical simulations were carried out on the Cray
XC50 at the Center for Computational Astrophysics, National
Astronomical Observatory of Japan.

\bibliographystyle{aasjournal}
\bibliography{references}

\begin{appendix}

\label{sec:appendix}

\section{Fit to the minimum mass}   \label{sec:fitminimum}

    The maximum value of $M_{\rm enc}/M_{\rm BE}$ increases with increasing mass of the star-forming halo. The mass ratio can be well approximated as $\log(M_{\rm enc}/M_{\rm BE})_{\rm max} = a (\log M)^2 +  b\log M + c$, where $a,b,c$ are fitting parameters. 
    We derive $M_{\rm min}$ by finding a root of $a (\log M)^2 +  b\log M + c = 0$ ($\Leftrightarrow \left[M_{\rm enc}/M_{\rm BE}\right]_{\rm max} = 1$) for each set of ($\metal,J_{21},\zin$). 
    By further fitting the derived $M_{\rm min}$ as a function of ($\metal,J_{21},\zin$),
    we obtain
    \[
    \log M_{\rm min} =  \frac{{ c_1(\zin)} + {c_2(\zin)} \log J_{21}}{1 + \tilde{Z}^{0.2}},
    \]
    where
    \gathering{
    c_1(\zin) = -0.080 (\zin + 1) +   7.4\nonumber,\\
    c_2(\zin) = -0.014 (\zin+1) +  0.36 \nonumber,\\
    \text{and}\quad \tilde{Z} \equiv \metal/\smetal. \nonumber
    }

\section{Parameters for gas Mass Evolution}    \label{sec:supplymental}
In \secref{sec:analytic}, 
we show an analytic model for the gas mass evolution of photoevaporating halos and also give relevant parameters. 
\tref{tab:data} lists those parameters for all of 495~runs. 
It also lists the resulting fitting parameters discussed in \secref{sec:evatime}.

\startlongtable
\begin{longrotatetable}
\begin{deluxetable*}{c C C C C  | C  C C CCCC|CCC }
  \tablecaption{ Parameters and resulting evaporation times of each run. \label{tab:data}}
    \hspace{-5cm}
  \tablehead{
Label	&
 {\metal}  	&    J_{21}   	& {M}   		&   {\zin}  		& 
 t_0		&
 q 			&   \eta 		&\chi	        &   \chi^{\prime}   & 
 \tilde{L}_{\rm s}	& \tilde{u}_{\rm IF}	&  C_1 	&
 \tilde{t}_{\rm s}	&	p	\\
 		&
  (\smetal)	&  (\Msun) 	&    			&   			& 
  (\Myr)  	&   
            &               &			    &		&			
                    &		                &       &			
            & (\Myr)
  }
  \startdata
  \simlabel{$\infty$}{0}{3}{10} & 0 & 1 & 10^{  3} & 10 & 76.9 & 14.9 &  119 & 30.7 & 29.5 & 1.36 & 0.444 & 0.00659\,(3.7\%) & 0.0228\,(0.12\%) &  3.7\,(0.38\%)\\ 
\simlabel{6}{0}{3}{10} & 10^{-6} & 1 & 10^{  3} & 10 & 76.9 & 14.9 &  119 & 30.7 & 29.5 & 1.36 & 0.444 & 0.00659\,(3.8\%) & 0.0228\,(0.12\%) &  3.7\,(0.38\%)\\ 
\simlabel{5}{0}{3}{10} & 10^{-5} & 1 & 10^{  3} & 10 & 76.9 & 14.9 &  119 & 30.7 & 29.5 & 1.36 & 0.444 & 0.00658\,(3.7\%) & 0.0228\,(0.12\%) &  3.7\,(0.38\%)\\ 
\simlabel{4}{0}{3}{10} & 10^{-4} & 1 & 10^{  3} & 10 & 76.9 & 14.9 &  119 & 30.7 & 29.5 & 1.36 & 0.444 & 0.00659\,(3.7\%) & 0.0228\,(0.12\%) &  3.7\,(0.38\%)\\ 
\simlabel{3}{0}{3}{10} & 10^{-3} & 1 & 10^{  3} & 10 & 76.9 & 14.9 &  119 & 30.7 & 29.5 & 1.36 & 0.444 & 0.00645\,(3.9\%) & 0.0228\,(0.12\%) & 3.67\,(0.38\%)\\ 
\simlabel{$\infty$}{1}{3}{10} & 0 & 0.1 & 10^{  3} & 10 & 76.9 & 1.49 & 11.9 & 9.71 & 9.34 & 0.136 & 4.44 & 1.94\e{-18}\,(-) & 0.0407\,(0.37\%) & 2.84\,(0.96\%)\\ 
\simlabel{6}{1}{3}{10} & 10^{-6} & 0.1 & 10^{  3} & 10 & 76.9 & 1.49 & 11.9 & 9.71 & 9.34 & 0.136 & 4.44 & 1.94\e{-18}\,(-) & 0.0407\,(0.37\%) & 2.84\,(0.96\%)\\ 
\simlabel{5}{1}{3}{10} & 10^{-5} & 0.1 & 10^{  3} & 10 & 76.9 & 1.49 & 11.9 & 9.71 & 9.34 & 0.136 & 4.44 & 1.94\e{-18}\,(-) & 0.0407\,(0.37\%) & 2.84\,(0.96\%)\\
\enddata
  \tablecomments{
  $t_{\rm lim}$ is the time at which halos satisfy the conditions of
  \eqnref{eq:limitingtime2}, 
  and $z_{\rm lim}$ is the corresponding redshift. 
  Hyphens in these columns indicate that halos have photoevaporated before the conditions. 
  Parentheses in the columns of the fitting parameters $C_1$, $\tilde{t}_{\rm s}$, and $p$ 
  show the fitting errors. 
  Hyphens in these columns mean that no fitting parameters are found with reasonable errors. 
  Note that $\tilde{t}_{\rm s}$ and $p$ are underivable when $C_1 = 1$.
  This table is published in its entirely in the machine-readable format. 
  }
\end{deluxetable*}
\end{longrotatetable}

\end{appendix}

\end{document}